\pdfoutput=1

\RequirePackage{ifpdf}
\documentclass[12pt,nohyper,twosided]{JHEP3}

\usepackage{epsfig}
\usepackage{float}
\usepackage{amsmath}
\usepackage{color}
\usepackage{array}
\let\normalcolor\relax
\usepackage{wick}
\usepackage{cite}
\usepackage{enumerate}
\usepackage{arydshln}

\newcommand{\be}{\begin{eqnarray}}
\newcommand{\ee}{\end{eqnarray}}

\newcommand{\smallminus}{{\rm\rule[2.4pt]{6pt}{0.65pt}}}
\newcommand{\smallplus}{\hspace{0.5pt}\text{{\small+}}\hspace{-0.5pt}}
\newcommand{\mi}{\smallminus}
\newcommand{\pl}{\smallplus}

   \newcommand{\la}{\langle}
\newcommand{\ra}{\rangle}

\title{\hspace{-0.0cm}{\LARGE The Amplituhedron}}
\author{\vspace{-.5cm}Nima Arkani-Hamed$^{a}$ and Jaroslav Trnka$^{b}$\\
{\footnotesize{\it $^{a}$ School of Natural Sciences, Institute for Advanced Study, Princeton, NJ 08540, USA}\\
{\it $^{b}$ California Institute of Technology, Pasadena, CA 91125,
USA}}\vspace{-.5cm}} \preprint{2013}

\abstract{Perturbative scattering amplitudes in gauge theories have
remarkable simplicity and hidden infinite dimensional symmetries
that are completely obscured in the conventional formulation of
field theory using Feynman diagrams. This suggests the existence of
a new understanding for scattering amplitudes where
locality and unitarity do not play a central role but are derived
consequences from a different starting point. In this note  we
provide such an understanding for ${\cal N} = 4$ SYM scattering
amplitudes  in the planar limit, which we identify as  ``the volume"
of a new mathematical object--the Amplituhedron--generalizing the
positive Grassmannian.  Locality and unitarity emerge hand-in-hand
from positive geometry.}

\preprint{CALT-68-2872}

\begin{document}

\newpage

\section{Scattering Without Space-Time}

Scattering amplitudes in gauge theories are amongst the most
fundamental observables in physics. The textbook approach to
computing these amplitudes in perturbation theory, using Feynman
diagrams, makes locality and unitarity as manifest as possible, at
the expense of introducing large amounts of gauge redundancy into
our description of the physics, leading to an explosion of apparent
complexity for the computation of amplitudes for all but the very
simplest processes. Over the last quarter-century it has become
clear that this complexity is a defect of the Feynman diagram
approach to this physics, and is not present in the final amplitudes
themselves, which are astonishingly simpler than indicated from the
diagrammatic expansion \cite{PT,Z1,Z2,Witten:2003nn, CSW,
BCFW1,BCFW2}.

This has been best understood  for maximally supersymmetric gauge
theories in the planar limit. Planar ${\cal N} = 4$ SYM has been
used as a toy model for real physics in many guises, but as toy
models go, its application to scattering amplitudes is closer to the
real world than any other. For instance the leading tree
approximation to scattering amplitudes is identical to ordinary
gluon scattering, and the most complicated part of loop amplitudes,
involving virtual gluons, is also the same in ${\cal N} = 4$ SYM as
in the real world.

Planar ${\cal N} = 4$ SYM amplitudes turn out to be especially
simple and beautiful, enjoying the hidden symmetry of dual superconformal
invariance\cite{DCI1,DCI2}, associated with a dual interpretation of
scattering amplitudes as a supersymmetric Wilson loop
\cite{WL1,WL2,Alday:2010zy}. Dual superconformal symmetry combines
with the ordinary conformal symmetry to generate an infinite
dimension ``Yangian" symmetry \cite{Yangian}. Feynman diagrams
conceal this marvelous structure precisely as a consequence of
making locality and unitarity manifest. For instance, the Yangian
symmetry is obscured in either one of the standard physical
descriptions either as a``scattering amplitude" in one space-time or
a ``Wilson-loop" in its dual.

This suggests that there must be a different formulation of the
physics, where locality and unitarity do not play a central role,
but emerge as derived features from a different starting point. A
program to find a reformulation along these lines was initiated in
\cite{N1,N2}, and  in the context of a planar ${\cal N} = 4$ SYM was
pursued in \cite{N3,N4,N5}, leading to a new physical and
mathematical understanding of scattering amplitudes \cite{N6}. This
picture builds on BCFW recursion relations for tree \cite{BCFW1,
BCFW2} and loop \cite{N5,Rutger} amplitudes, and represents the
amplitude as a sum over basic building blocks, which can be
physically described as arising from gluing together the elementary
three-particle amplitudes to build more complicated on-shell
processes. These ``on-shell diagrams" (which are essentially the
same as the ``twistor diagrams" of \cite{TD1,TD2,N3}) are remarkably
connected with ``cells" of a beautiful new structure in algebraic
geometry, that has been studied by mathematicians over the past
number of years, known as the positive Grassmannian \cite{alex, N6}.
The on-shell building blocks can not be associated with  local
space-time processes. Instead, they enjoy all the symmetries of the
theory, as made manifest by their connection with the
Grassmannian--indeed, the infinite dimensional Yangian symmetry is
easily seen to follow from ``positive" diffeomorphisms \cite{N6}.

While these developments give a complete understanding for the
on-shell building blocks of the amplitude, they do not go further to
explain {\it why} the building blocks have to be combined in a
particular way to determine the full amplitude itself. Indeed, the
particular combination of on-shell diagrams is dictated by {\it
imposing} that the final result is local and unitary--locality and
unitarity specify the singularity structure of the amplitude, and
this information is {\it used} to determine the full integrand. This
is unsatisfying, since we want to see locality and unitarity emerge
from more primitive ideas, not merely use them to obtain the
amplitude.

An important clue \cite{N4,A1,N6} pointing towards a deeper understanding is that
the on-shell representation of scattering
amplitudes is not unique: the recursion relations can be solved in
many different ways, and so the final amplitude can be expressed as
a sum of on-shell processes in different ways as well. The on-shell
diagrams satisfy remarkable identities--now  most easily understood
from their association with cells of the positive Grassmannian--that
can be used to establish these equivalences. This observation led
Hodges \cite{A1} to a remarkable observation for the simplest case of
``NMHV" tree amplitudes, further developed  in \cite{N7}: the amplitude
can be thought of as the volume of a certain polytope in momentum
twistor space. However there was no a priori
understanding of the origin of this polytope, and the picture
resisted a direct generalization to more general trees or to loop amplitudes.
Nonetheless, the polytope idea motivated a continuing search for a geometric representation of the amplitude
 as ``the volume" of ``some canonical region" in
``some space", somehow related to the positive Grassmannian, with
different ``triangulations" of the space corresponding to different
natural decompositions of the amplitude into building blocks.

In this note we finally realize this picture. We will introduce a
new mathematical object whose ``volume" directly computes the
scattering amplitude. We call this object the ``Amplituhedron", to
denote its connection both to scattering amplitudes and positive
geometry. The amplituhedron can be given a self-contained definition
in a few lines as done below in section 9. We will motivate its
definition from elementary considerations,  and show how scattering
amplitudes are extracted from it.

Everything flows from generalizing the notion of the ``inside of a
triangle in a plane". The first obvious generalization is
to the inside of a simplex in projective space, which further
extends to the positive Grassmannian. The second generalization
is to move from triangles to convex polygons, and then extend this
into the Grassmannian. This gives us the amplituhedron for tree
amplitudes, generalizing the positive Grassmannian by extending the
notion of positivity to include external kinematical data. The full
amplituhedron at all loop order further generalizes the notion of
positivity in a way motivated by the natural idea of ``hiding
particles".

Another familiar  notion associated with triangles and polygons is
their area. This is more naturally described in a projective way by
a canonical 2-form with logarithmic singularities on the boundaries
of the polygon. This form also generalizes to the full
amplituhedron, and determines the (integrand of) the scattering
amplitude. The geometry of the amplituhedron is completely bosonic,
so the extraction of the superamplitude from this canonical form
involves a novel treatment of supersymmetry, directly motivated by
the Grassmannian structure.

The connection between the amplituhedron and scattering amplitudes
is a conjecture which has  passed a large number of non-trivial
checks, including an understanding of how locality and unitarity
arise as consequences of positivity. Our purpose in this note is to
motivate and give the complete definition of the  amplituhedron and
its connection to the superamplitude in planar ${\cal N}=4$ SYM. The
discussion will be otherwise telegraphic and few details or examples
will be given. In two accompanying notes \cite{Into, Threeviews}, we
will initiate a systematic exploration of various aspects of the
associated geometry and physics. A much more thorough exposition of
these ideas, together with many examples worked out in detail, will
be presented in \cite{Long}.

\subsection*{Notation}

The external data for massless $n$ particle scattering amplitudes
(for an excellent review see \cite{review}) are labeled as
$|\lambda_a,\tilde \lambda_a, \tilde \eta_a \rangle$ for $a=1,
\dots, n$. Here $\lambda_a, \tilde \lambda_a$ are the
spinor-helicity variables, determining null momenta $p_a^{A \dot{A}}
= \lambda_a^A \tilde \lambda_a^{\dot{A}}$. The $\tilde \eta_a$ are
(four) grassmann variables for  on-shell superspace. The component
of the color-stripped superamplitude with weight $4(k+2)$ in the $\tilde \eta$'s is
$M_{n,k}$. We can write \be M_{n,k}(\lambda_a, \tilde \lambda_a,
\tilde \eta_a) = \frac{\delta^4(\sum_a \lambda_a \tilde \lambda_a)
\delta^8(\sum_a \lambda_a \tilde \eta_a)}{\langle 1 2 \rangle \dots
\langle n 1 \rangle} \times {\cal M}_{n,k}( z_a, \eta_a) \ee where
$(z_a, \eta_a)$ are the (super) ``momentum-twistor" variables
\cite{A1}, with $ z_a = \left(\begin{array}{c} \lambda_a \\ \mu_a
\end{array} \right)$. The $z_a, \eta_a$ are unconstrained, and
determine the $\lambda_a, \tilde \lambda_a$ as
\begin{eqnarray}
\tilde \lambda_a &=& \frac{\langle a\mi1 \, a \rangle \mu_{a\pl1} + \langle a\pl1 \, a\mi1
\rangle \mu_a + \langle a \, a\pl1 \rangle \mu_{a\mi1}}{\langle a\mi1 \, a \rangle \langle a \, a\pl1 \rangle}, \nonumber \\
\tilde \eta_a &=& \frac{\langle a\mi1 \, a \rangle \eta_{a\pl1} +
\langle a\pl1 \, a\mi1 \rangle \eta_a + \langle a \, a\pl1 \rangle
\eta_{a\mi1}}{\langle a\mi1 \, a \rangle \langle a \, a\pl1 \rangle}
\end{eqnarray}
where throughout
this paper, the angle brackets $\langle \dots \rangle$ denotes
totally antisymmetric contraction with an $\epsilon$ tensor.
${\cal M}_{n,k}$ is cyclically invariant. It is also invariant under the little group action $(z_a, \eta_a)
\to t_a (z_a, \eta_a)$, so $(z_a, \eta_a)$ can be taken to live in
$\mathbb{P}^{3|4}$.

At loop level, there is a well-defined notion of ``the integrand"
for scattering amplitudes, which at $L$ loops is a $4L$ form. The
loop integration variables are points in the (dual) spacetime $x^\mu_i$,
which in turn can be associated with $L$ lines in momentum-twistor
space that we denote as ${\cal L}_{(i)}$ for $i = 1, \cdots, L$. The
$4L$ form is \cite{Andrewloop, LDbox, LocalIntegrand} \be {\cal
M}(z_a, \eta_a; {\cal L}_{(i)}) \ee We can specify the line by
giving two points ${\cal L}_{1 (i)},{\cal L}_{2 (i)}$ on it, which
we can collect as ${\cal L}_{\gamma(i)}$ for $\gamma = 1,2$. ${\cal
L}$ can also be thought of as a $2$ plane in 4 dimensions. In
previous work, we have often referred to the two points on the line
${\cal L}_1, {\cal L}_2$ as ``$AB$", and we will use this notation here
as well.

Dual superconformal symmetry says that ${\cal M}_{n,k}$ is invariant
under the $SL(4|4)$ symmetry acting on $(z_a,\eta_a)$ as
(super)linear transformations. The full symmetry of the theory is the Yangian of
$SL(4|4)$.

\section{Triangles $\to$ Positive Grassmannian}

To begin with, let us start with the simplest and most familiar
geometric object of all, a triangle in two dimensions. Thinking
projectively, the vertices are $Z_1^I,Z_2^I,Z_3^I$ where $I=1,\dots,
3$. The interior of the triangle is a collection of points of the
form \be Y^I = c_1 Z_1^I + c_2 Z_2^I + c_3 Z_3^I \ee
where we span over all $c_a$ with
\be
c_a > 0
\ee
$$
\includegraphics[scale=.75]{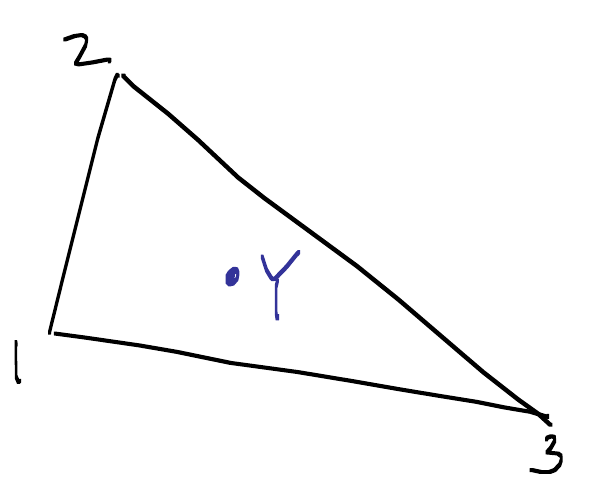}
$$
More precisely, the interior of a triangle is associated with a
triplet $(c_1,c_2,c_3)/GL(1)$, with all ratios $c_a/c_b > 0$, so
that the $c_a$ are either all positive or all negative, but here and
in the generalizations that follow,  we  will abbreviate this by
calling them all positive.  Including the closure of the triangle
replaces ``positivity"  with ``non-negativity", but we will continue
to refer to this as ``positivity" for brevity.

One obvious generalization
of the triangle is to an $(n-1)$ dimensional simplex in a general
projective space, a collection $(c_1, \dots, c_n)/GL(1)$, with $c_a
> 0$. The $n$-tuple $(c_1, \dots, c_n)/GL(1)$ specifies a line in $n$
dimensions, or a point in $\mathbb{P}^{n-1}$. We can generalize this
to the space of $k$-planes in $n$ dimensions--the Grassmannian
$G(k,n)$--which we can take to be a collection of $n$
$k-$dimensional vectors modulo $GL(k)$ transformations, grouped into
a $k \times n$ matrix \be C =   \left( \begin{array}{ccc} & & \\ c_1
&  \dots &  c_n \ \\ & & \end{array} \right)/GL(k) \ee

Projective space is the special case of $G(1,n)$. The notion of
positivity giving us the ``inside of a simplex" in projective space
can be generalized to the Grassmannian. The only possible $GL(k)$
invariant notion of positivity for the matrix $C$ requires us to fix
a particular ordering of the columns, and demand that all minors in
this ordering are positive: \be \langle c_{a_1} \dots c_{a_k}
\rangle > 0 \, \, {\rm for} \, \, a_1<\dots<a_k \ee We can also talk
about the very closely related space  of positive matrices
$M_+(k,n)$, which are just $k \times n$ matrices with all positive
ordered minors. The only difference with the positive Grassmannian
is that in $M_+(k,n)$ we are not moding out by $GL(k)$.

Note that while  both $M_+(k,n)$ and $G_+(k,n)$ are defined with a
given ordering for the columns of the matrices, they have a natural
cyclic structure. Indeed, if $(c_1, \dots, c_n)$ give a positive
matrix, then positivity is preserved under the  (twisted) cyclic
action  $c_1 \to c_2, \dots, c_n \to (-1)^{k-1} c_1$.

\section{Polygons $\to$ (Tree) Amplituhedron ${\cal A}_{n,k}(Z)$}

Another natural generalization of a triangle is to a more
general polygon with $n$ vertices $Z^I_1, \dots, Z^I_n$. Once again
we would like to discuss the interior of this region. However in
general this is not canonically defined--if the points $Z_a$ are
distributed randomly, they don't obviously enclose a region where all
the $Z_a$ are all vertices. Only if the $Z_a$ form a {\it convex}
polygon do we have a canonical ``interior" to talk about:

$$
\includegraphics[scale=.65]{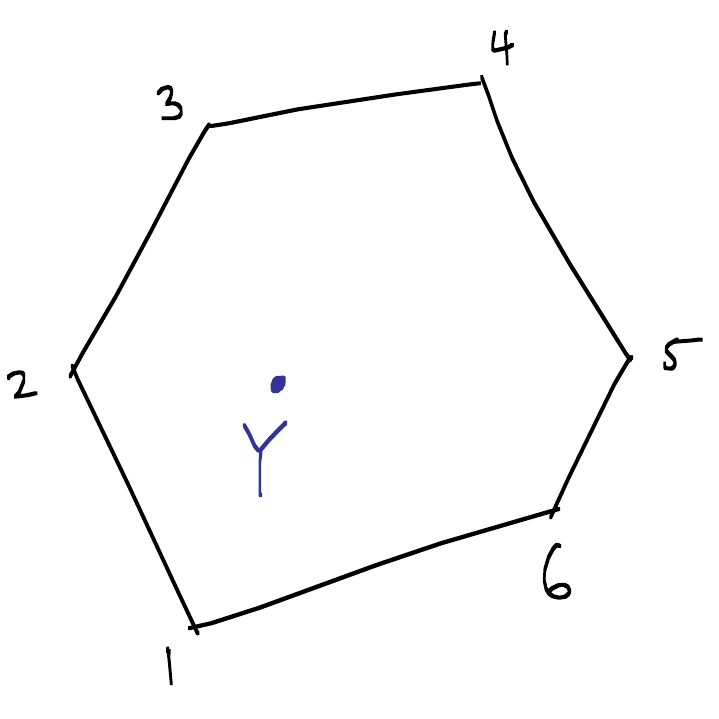}
$$

Now,  convexity for the $Z_a$ is a special case of positivity in the
sense of the positive matrices we have just defined. The points
$Z_a$ form a closed polygon only if the $3 \times n$ matrix with
columns $Z_a$ has all positive (ordered) minors: \be \langle Z_{a_1}
Z_{a_2} Z_{a_3} \rangle > 0 \quad {\rm for} \quad a_1 < a_2 < a_3
\ee Having arranged for this, the interior of the polygon is given
by points of the form \be Y^I = c_1 Z_1^I + c_2 Z_2^I + \dots c_n
Z_n^I \quad {\rm with}\quad  c_a > 0 \ee Note that this can be
thought of as an interesting pairing of two different positive
spaces. We have \be (c_1, \dots, c_n) \subset G_+(1,n),\quad
\left(Z_1, \dots, Z_n \right) \subset M_+(3,n) \ee If we jam them
together to produce \be Y^I = c_a Z_a^I \ee for fixed $Z_a$, ranging
over all $c_a$ gives  us all the points on the inside of the
polygon, living in $G(1,3)$.

This object has a natural generalization to higher projective
spaces; we can consider $n$ points $Z_a^I$ in $G(1,1+m)$, with $I =
1, \dots, 1+m$, which are positive \be \langle Z_{a_1} \dots Z_{a_{1
+ m}} \rangle > 0 \ee Then, the analog of the ``inside of the
polygon" are points of the form \be Y^I = c_a Z_a^I, \quad {\rm
with} \quad c_a > 0 \ee This space is very closely related to the
``cyclic polytope" \cite{cyclic}, which is the convex hull of $n$
ordered points on the moment curve in $\mathbb{P}^{m}$, $Z_a =
(1,t_a,t_a^2, \dots, t_a^{m})$, with $t_1 < t_2 \dots < t_n$. From
our point of view, forcing the points to lie on the moment curve is
overly restrictive; this is just one way of ensuring the positivity
of the $Z_a$.

We can further generalize this structure into the Grassmannian. We
take positive  external data as $(k + m)$ dimensional vectors
$Z_a^I$ for $I =1, \dots, k+m$.  It is natural to restrict $n \geq
(k+m)$, so that the external $Z_a$ fill out the entire $(k+m)$
dimensional space. Consider the space of $k$-planes in this $(k+m)$
dimensional space, $Y \subset G(k,k+m)$, with co-ordinates \be
Y_\alpha^I, \, \alpha = 1, \dots k, \, I = 1, \dots, k+m \ee We then
consider a subspace of $G(k,k+m)$ determined by taking all possible
``positive" linear combinations of the external data, \be Y = C
\cdot Z \ee or more explicitly \be Y_\alpha^I = C_{\alpha a} Z_a^I
\ee where \be C_{\alpha a} \subset G_+(k,n), Z_a^I \subset
M_+(k+m,n) \ee
It is trivial to see that this space is cyclically invariant if $m$ is even: under the twisted cyclic symmetry,
$Z_n \to (-1)^{k+m-1} Z_1$ and $c_n \to (-1)^{k-1} c_1$, and the product is invariant for even $m$.

We call this space the generalized tree amplituhedron ${\cal A}_{n,k,m}(Z)$. The polygon is the simplest case with $k=1,m=2$. Another special case is $n = (k+m)$, where we can  use $GL(k+m)$
transformations to set the external data to the identity matrix
$Z_a^I = \delta_a^I$. In this case ${\cal A}_{k+m,k,m}$ is identical to the usual
positive Grassmannian $G_+(k,k+m)$.

The case of immediate relevance to physics is $m=4$, and we will refer to this as the tree amplituhedron
${\cal A}_{n,k}(Z)$. The tree amplituhedron lives in a
$4k$ dimensional space and is not trivially visualizable. For $k=1$, it
is  a polytope, with inequalities determined by linear
equations, while for $k>1$, it is not a polytope and is
more ``curvy". Just to have a picture,
below we sketch a 3-dimensional face of the 4 dimensional
amplituhedron for $n=8$, which turns out to be the space $Y = c_1 Z_1 +\dots c_7
Z_7$ for $Z_a$ positive external data  in $\mathbb{P}^3$:
$$
\includegraphics[scale=.6]{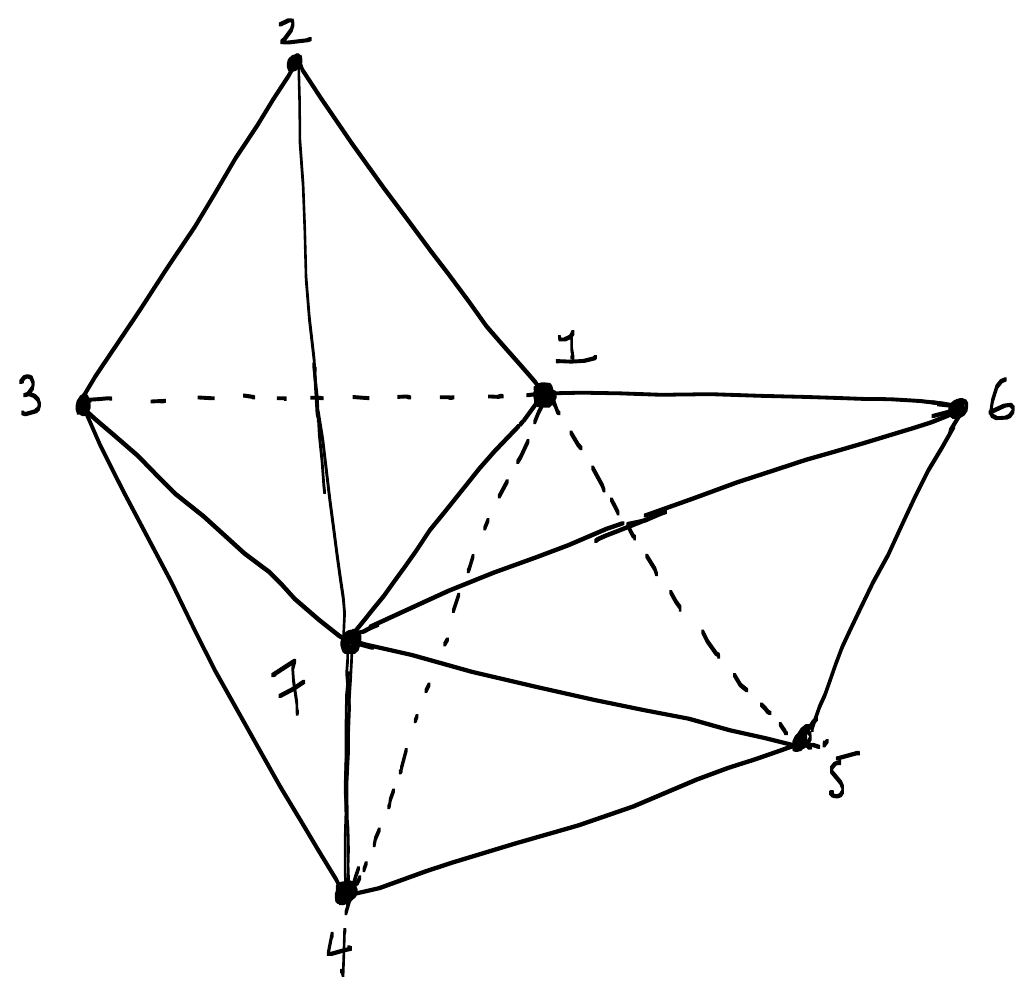}
$$

\section{Why Positivity?}

We have motivated the structure of the amplituhedron by mimicking
the geometric idea of the ``inside" of a convex polygon.
However there is a simpler and deeper origin of  the need for
positivity. We can  attempt to define $Y = C \cdot Z$ with no
positive  restrictions on $C$ or $Z$. But in general, this will not
be projectively meaningful, and this expression  won't allow us to
define a region in $G(k,k+m)$. The reason is that  for $n > k+m$,
there is always some linear combination of the $Z_a$ which sum to
zero! We have to take care to avoid this happening, and in order to
avoid ``0" on the left hand side, we obviously need positivity
properties on both the $Z$'s and the $C$'s.

It is simple and instructive to see why positivity ensures that the
$Y = C \cdot Z$ map is projectively well-defined. We will see this
as a by-product of locating  the co-dimension one boundaries of the
generalized tree amplituhedron. Let us illustrate the idea already
for the simplest case of the polygon with $k=1, m=2$, with $Y = c_1
Z_1 + \dots c_n Z_n$.  In order to look at the boundaries of the
space, let us compute $\langle Y Z_i Z_j \rangle$ for some $i,j$. If
as we sweep through all the allowed $c$'s, $\langle Y Z_i Z_j
\rangle$ changes sign from being positive to negative, then
somewhere $\langle Y Z_i Z_j \rangle \to 0$ and $Y$ lies on the line
$(Z_i Z_j)$ in the interior of the space, thus $(Z_i Z_j)$ should
not be a boundary of the polygon. On the other hand, if $\langle Y
Z_i Z_j \rangle$ everywhere has a uniform sign, then $(Z_i Z_j)$ is
a boundary of the polygon:

$$
\includegraphics[scale=.6]{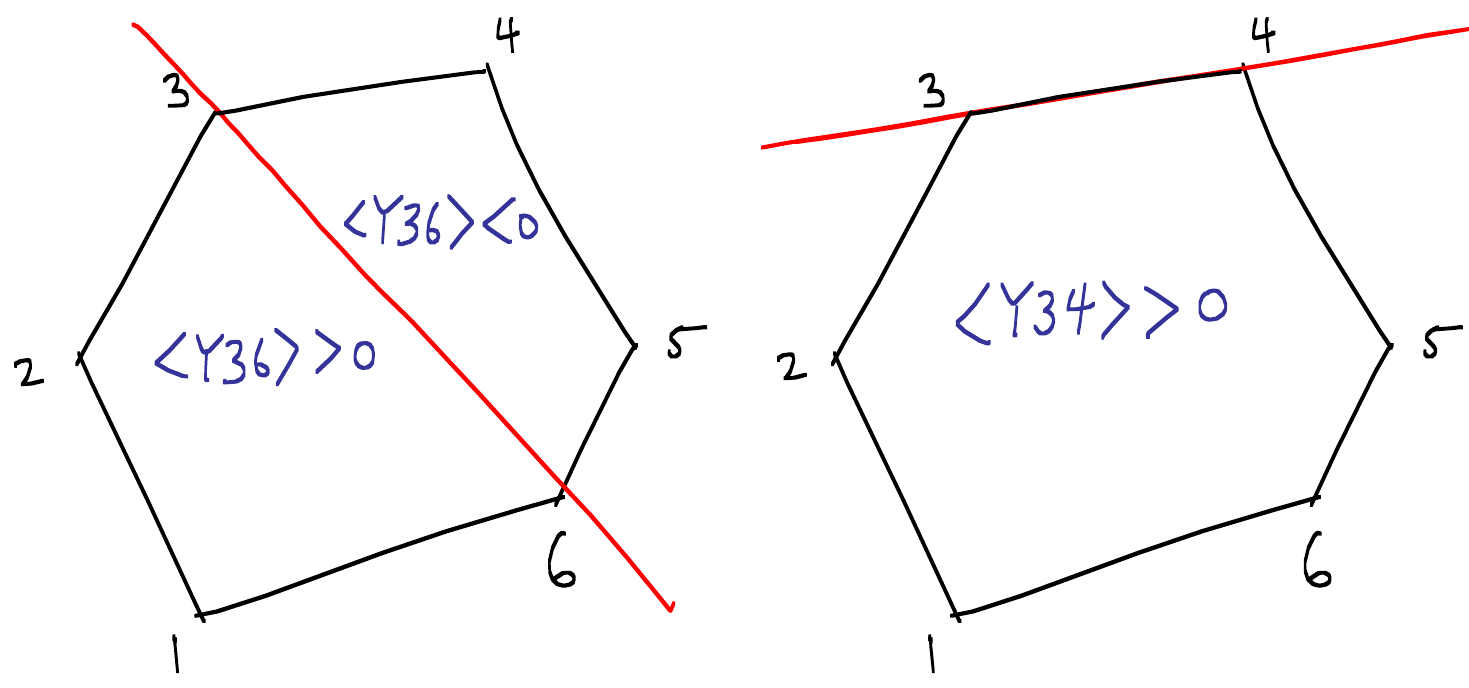}
$$

Of course for the polygon it is  trivial to directly see that the
co-dimension one boundaries are just the lines $(Z_i Z_{i+1})$, but
we wish to see this more algebraically, in a way that will
generalize to the amplituhedron where ``seeing" is harder.
So, we compute \be \langle Y Z_i Z_j \rangle =\sum_a c_a \langle Z_a
Z_i Z_j \rangle \ee We can see why there is some hope for the
positivity of this sum, since the $c_a > 0$, and also ordered minors
of the $Z's$ are positive. It is however obvious that if $i,j$ are
not consecutive, some of the terms in this sum will be positive, but
some (where $a$ is stuck between $i,j$) will be negative. But
precisely when $i,j$ are consecutive, we get a manifestly
positive sum: \be \langle Y Z_i Z_{i+1} \rangle = \sum_a c_a \langle
Z_a Z_i Z_{i+1} \rangle > 0 \ee Since $\langle Z_a Z_i Z_{i+1}
\rangle > 0$ for $a \neq i, i+1$, this is manifestly positive. Thus
the boundaries are lines $(Z_i Z_{i+1})$ as expected.

This also tells us that the map  $Y = C \cdot Z$ is projectively well-defined.
There is no way to get $Y \to 0$, since this would make the left hand side
identically zero, which is impossible without making all the $c_a$
vanish, which is not permitted as we we mod out by
$GL(1)$ on the $c_a$.

We can extend this logic to higher $k,m$. Let's look at the case
$m=4$ already for $k=1$.
We can investigate whether the plane $(Z_i Z_j Z_k Z_l)$ is a boundary by computing
\be \langle Y Z_i Z_j Z_k Z_l \rangle = \sum_a c_a  \langle Z_a Z_i
Z_j Z_k Z_l \rangle \ee Again, this is not in general positive. Only for
$(i,j,k,l)$ of the form $(i,i+1,j, j+1)$, we have \be \langle Y Z_i
Z_{i+1} Z_j Z_{j+1} \rangle = \sum_a c_a  \langle Z_a Z_i Z_{ i+1}
Z_j Z_{j+1} \rangle > 0 \ee For general even the  $m$, the
boundaries are when $Y$ is on the plane\\ $(Z_{i} Z_{i+1}
\dots Z_{i_{m/2 - 1}} Z_{i_{m/2}})$. This again shows that the $Y =
C \cdot Z$ is projectively well-defined. The result extends
trivially to general $k$, provided the positivity of $C$ is
respected. For $m = 4$ the boundaries are again when the $k$-plane $(Y_1 \cdots Y_k)$ is on  $(Z_i Z_{i+1} Z_j
Z_{j+1})$, as follows from \be \langle Y_1 \dots Y_k Z_i Z_{i+1} Z_j
Z_{j+1} \rangle = \sum_{a_1< \dots < a_k} \langle c_{a_1} \dots
c_{a_k} \rangle  \langle Z_{a_1} \dots Z_{a_k} Z_i Z_{i+1} Z_j
Z_{j+1} \rangle > 0 \ee which also shows that $Y$ is always a full rank
$k$-plane in $k+4$ dimensions.

The emergence of boundaries on the plane $(Z_{i} Z_{i+1} Z_j
Z_{j+1})$ is a simple and striking consequence of positivity. We will
shortly understand that the location of these boundaries are the
``positive origin" of locality from the geometry of the
amplituhedron.

\section{Cell Decomposition}

The tree amplituhedron can be thought of as the image of the
top-cell of the the positive Grassmannian $G_+(k,n)$ under the map
$Y = C \cdot Z$. Since ${\rm dim}\, G(k,k+m) = m k \leq\,{\rm
dim}\,G(k,n)= k (n -k)$ for $n \geq k+m$, this is in general a
highly redundant map. We can already see this in the simplest case
of the polygon, which lives in 2 dimensions, while the $c_a$ span an
$(n-1)$ dimensional space. The non-redundant maps into $G(k,k+m)$
can only come from the $m \times k$ dimensional ``cells" of
$G_+(k,n)$. For the polygon, these are the cells we can label as
$(i,j,k)$, where all but $(c_i,c_j,c_k)$ are non-vanishing. The
image of these cells in the $Y$-space are of course just the
triangles with vertices at $Z_i,Z_j,Z_k$, which lie inside the
polygon.

The  union of all these triangles covers the inside of the polygon.
However, we can also cover the inside of the polyon more nicely with
non-overlapping triangles, giving a triangulation. Said in a
heavy-handed way, we find a collection of 2 dimensional cells of
$G_+(1,n)$, so that their images in $Y$ space are non-overlapping
except on boundaries, and collectively cover the entire polygon. Of
course these collections of cells are not unique--there are many
different triangulations of the polygon. A particularly simple one
is
$$
\includegraphics[scale=.6]{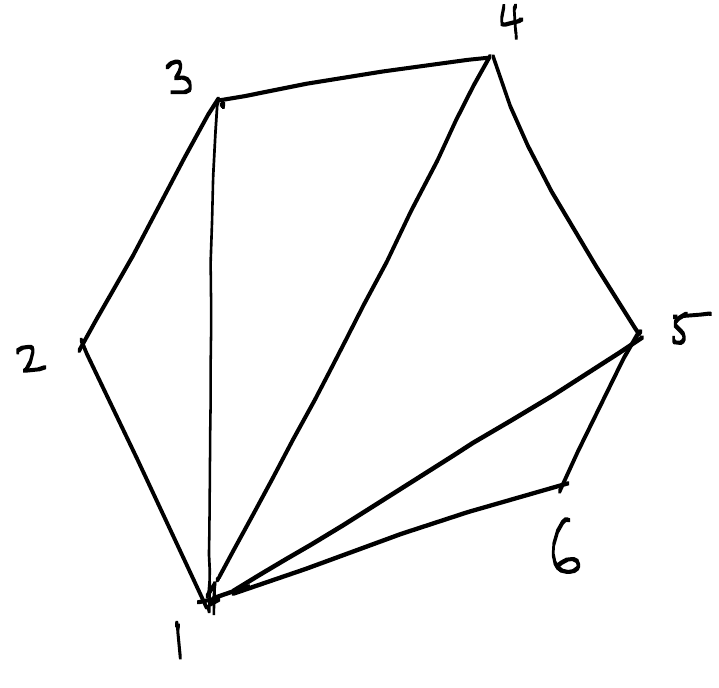}
$$
which we can write as \be \sum_i (1\,i\,i\pl1) \ee Sticking with
$k=1$ but moving to $m=4$, the four-dimensional cells of $G_+(1,n)$
are labeled by five non-vanishing $c$'s  $(c_i,c_j,c_k,c_l,c_m)$.
While it is harder to visualize, one can easily show algebraically
that the above simple triangulation of the polygon generalizes to
\be \sum_{i<j} (1\,i\,i\pl1\,j\,j\pl1) \ee

This expression is immediately recognizable  to physicists familiar
with scattering amplitudes in ${\cal N} = 4$ SYM. If  the
$(i,j,k,l,m)$ are interpreted as ``R-invariants", this expression is
one of the canonical BCFW representations of the $k=1$ ``NMHV" tree
amplitudes. In the positive Grassmannian representation for
amplitudes \cite{N4,N6},  R-invariants are precisely associated with
the corresponding four-dimensional cells of $G(1,n)$.

For general $k$, $m$ any $m \times k$ dimensional cell of $G_+(k,n)$
maps under $Y = C \cdot Z$ into some region or cell in $G(k,k+m)$.
Said more explicitly, consider an $m \times k$ dimensional cell
$\Gamma$ of the $G_+(k,n)$, with ``positive co-ordinates"
$C^\Gamma(\alpha^\Gamma_1, \dots, \alpha^\Gamma_{m \times k})$ \cite{N6}.
Putting $Y = C(\alpha) \cdot Z$ and scanning over all positive
$\alpha$'s, this carves out a region in $G(k,k+m)$ which is a
corresponding cell $\Gamma$ of the tree amplituhedron. A cell
decomposition is a collection $T$ of non-overlapping cells $\Gamma$
which cover the entire amplituhedron.

The case of immediate relevance for physics is $m=4$. For any $k$,
the BCFW decomposition of tree amplitudes gives us a collection of
$4 \times k$ dimensional cells of the positive Grassmannian. We have
performed extensive checks for high $k$ and $n$, that for positive
external $Z$, under $Y = C \cdot Z$ these cells are beautifully
mapped into non-overlapping regions of $G(k,k+4)$ that together
cover the entire tree amplituhedron. As we have stressed, other than
the desire to make the final result local and unitary, we did not
previously have a rational for thinking about this particular
collection of cells of $G_+(k,n)$. Now we know what natural question
this collection of cells are answering: taken together they
``cellulate" the tree amplituhedron.  We will shortly see how to
directly associate the amplitude itself directly with the geometry
of the amplituhedron.

\section{``Volume" as Canonical Form}

Before discussing how to determine the (super)amplitude from the
geometry, let us define the notion of a ``volume" associated with
the amplituhedron. As should by now be expected, we will merely generalize a simple
existing idea from the world of triangles and polygons.

The usual notion of ``area" has units and is obviously not projectively
meaningful. However there is a closely related idea that is. For the
triangle, we can consider a rational 2-form in $Y$-space,  which has
logarithmic singularities on the boundaries of the triangle. This is
naturally associated with positive co-ordinates for the triangle, if
we expand $Y = Z_3 + \alpha_1 Z_1 + \alpha_2 Z_2$, then the form is \be
\Omega_{123} = \frac{d\alpha_1}{\alpha_1} \frac{d\alpha_2}{\alpha_2} \ee which can be
re-written more invariantly as \be \Omega_{123} = \frac{\langle Y d
Y d Y \rangle \langle 1 2 3 \rangle^2}{\langle Y 12 \rangle \langle
Y 2 3 \rangle \langle Y 3 1 \rangle} \ee We can extend this to a
form $\Omega_P$ for the convex polygon $P$. The defining property of
$\Omega_P$ is that

\begin{center}
$\Omega_P$ has logarithmic singularities on all the boundaries of
$P$.
\end{center}
$\Omega_P$ can be obtained by first triangulating the polygon in
some way, then summing the elementary two-form for each triangle,
for instance as \be \Omega_P = \sum_i \Omega_{1\,i\,i\pl1}. \ee Each
term in this sum has singularities corresponding to $Y$ hitting the
boundaries of the corresponding triangle. Most of these singularities, associated
with the internal edges of the triangulation, are spurious, and
cancel in the sum. Of course the full form $\Omega_P$ is independent
of the particular triangulation.

This form is closely related to an area, not directly of the polygon
$P$, but its dual $\tilde{P}$, which is also a convex polygon
\cite{N7}. If we dualize so that points are mapped to lines and
lines to points, then a point $Y$ {\it inside} $P$ is mapped to a
line $Y$ {\it outside} $\tilde{P}$. If we write $\Omega_P = \langle
Y d^2 Y \rangle V(Y)$, then $V(Y)$ is the area of $\tilde{P}$ living
in the euclidean space defined by $Y$ as the line at infinity.

This form can be generalized to the tree amplituhedron in an obvious
way. We define a rational form $\Omega_{n,k}(Y;Z)$ with the property
that

\begin{center}
$\Omega_{n,k}(Y;Z)$ has logarithmic singularities on all the
boundaries of ${\cal A}_{n,k}(Z)$.
\end{center}

Just as for the polygon, one concrete way of computing  $\Omega$ is
to give a cell decomposition of the amplituhedron. For any cell
$\Gamma$ associated with positive co-ordinate $(\alpha^\Gamma_1,
\dots, \alpha^\Gamma_{4k})$, there is an associated form with
logarithmic singularities on the boundaries of the cell \be
\Omega_{n,k}^\Gamma(Y; Z) = \prod_{i = 1}^{4k} \frac{d
\alpha_i^\Gamma}{\alpha_i^\Gamma} \ee For instance, consider 4
dimensional cells for $k=1$, associated with cells in $G_+(1,n)$
which are vanishing for all but columns $a_1, \dots, a_5$, with
positive co-ordinates\\  $(\alpha_{a_1}, \dots, \alpha_{a4},
\alpha_{a_5} =1)$.  Its image in $Y$ space is simply \be Y =
\alpha_{a1} Z_{a_1} + \dots \alpha_{a_4} Z_{a_4}  + Z_{a_5} \ee and
the form is \be \frac{d \alpha_{a1}}{\alpha_{a1}} \dots \frac{d
\alpha_{a4}}{\alpha_{a4}} = \frac{\langle Y d^4 Y \rangle \langle
Z_{a_1} Z_{a_2} Z_{a_3} Z_{a_4} Z_{a_5} \rangle^4}{\langle Y Z_{a_1}
Z_{a_2} Z_{a_3} Z_{a_4} \rangle \dots \langle Y Z_{a_5} Z_{a_1}
Z_{a_2} Z_{a_3} \rangle} \ee Now, given a collection of cells $T$
that cover the full amplituhedron, $\Omega_{n,k}(Y;Z)$ is given by
\be \Omega_{n,k}(Y;Z) = \sum_{\Gamma \subset T}
\Omega_{n,k}^\Gamma(Y;Z) \ee As with the polygon, the form is
independent of the particular cell decomposition.

Note that the definition of the amplituhedron itself crucially
depends on the positivity of the external data $Z$, and this
geometry in turn determines the form $\Omega$. However, once this
form is in hand, it can be analytically continued to any
(complex!) $Y$ and $Z$.

\section{The Superamplitude}

We have already defined central objects in our story: the
tree amplituhedron, together with the associated form $\Omega$
that is loosely speaking its ``volume". The
tree super-amplitude ${\cal M}_{n,k}$ can be
directly extracted from $\Omega_{n,k}(Z)$. To see how, note that
we we can always use a $GL(4 +k)$ transformation to send $Y \to Y_0$
as \be Y_0 = \left(\begin{array}{ccc} & 0_{4 \times k} & \\
\hdashline & 1_{k \times k} & \end{array} \right) \ee We can think
of the 4 dimensional space complementary to $Y_0$, acted on by an
unbroken $GL(4)$ symmetry, as the usual $\mathbb{P}^3$ of
momentum-twistor space. Accordingly, we identify the top four
components of the $Z_a$ with the usual bosonic momentum-twistor
variables $z_a$: \be Z_a = \left(\begin{array}{c} z_a \\ *_1 \\
\vdots \\ *_k
\end{array}\right) \ee We still have to decide how to interpret the
remaining $k$ entries of $Z_a$. Clearly, if they are normal bosonic
variables, we have an infinite number of extra degrees of freedom.
It is therefore natural to try and make the remaining components
infinitesimal, by saying that some ${\cal N} + 1$'st power of them
vanishes. This is equivalent to saying that each entry can be
written as \be Z_a = \left(\begin{array}{c} z_a \\ \phi^A_1 \cdot
\eta_{1 A} \\ \vdots \\ \phi^A_k \cdot \eta_{A k} \end{array}\right)
\ee where $\phi_{1,\dots,k}$ and $\eta_a$ are Grassmann parameters,
and $A = 1, \dots, {\cal N}$.

Now there is a unique way to extract the amplitude. We simply
localize the form $\Omega_{n,k}(Y;Z)$ to $Y_0$, and integrate over the
$\phi$'s: \be {\cal M}_{n,k}(z_a,\eta_a) = \int d^{\cal N} \phi_1 \dots
d^{\cal N} \phi_k \int \Omega_{n,k}(Y;Z)  \delta^{4k}(Y;Y_0)
\label{super}\ee Here $\delta^{4k}(Y;Y_0)$ is a projective $\delta$
function \be \delta^{4k}(Y;Y_0) = \int d^{k \times k}
\rho_\alpha^\beta \, {\rm det}(\rho)^4 \, \delta^{k \times
(k+4)}(Y_\alpha^I - \rho_{\alpha}^\beta Y_{0 \beta}^I) \ee Note that
there is really no integral to perform in the second step; the delta
functions fully fix $Y$. Any form on $G(k, k+4)$ is of the form
\begin{equation}
\Omega =  \langle Y_1 \dots Y_k d^4 Y_1 \rangle \dots \langle Y_1
\dots Y_k d^4 Y_k \rangle \times \omega_{n,k}(Y;Z)
\end{equation}
and our expression just says that \be {\cal M}_{n,k}(z_a,\eta_a) = \int
d^{\cal N} \phi_1 \dots d^{\cal N} \phi_k  \omega_{n,k}(Y_0;Z_a) \ee
Note that we can define this operation for any ${\cal N}$, however,
only for ${\cal N} = 4$ does ${\cal M}_{n,k}$ further have weight zero under the
rescaling $(z_a,\eta_a)$.

This connection between the form and the super-amplitude also allows
us to directly exhibit the relation between our super-amplitude
expressions and the Grassmannian formulae of \cite{N4,N6}. Consider
the form in $Y$-space associated with a given $4k$ dimensional cell
$\Gamma$ of $G_+(k,n)$. Then, if $C^\Gamma_{\alpha
a}(\alpha_1,\dots, \alpha_{4k})$ are positive co-ordinates for the
cell, and $\Omega^\Gamma = \frac{d\alpha^\Gamma_1}{\alpha^\Gamma_1}
\dots \frac{d\alpha^\Gamma_{4k}}{\alpha^\Gamma_{4k}}$ is the
associated form in $Y$ space, then it is easy to show that \be \int
d^4 \phi_1 \dots d^4 \phi_k \int \Omega^\Gamma \delta^{4k}(Y;Y_0) =
\int \frac{d\alpha^\Gamma_1}{\alpha^\Gamma_1} \dots
\frac{d\alpha^\Gamma_{4k}}{\alpha^\Gamma_{4k}}
\delta^{4k|4k}(C_{\alpha a}(z) {\cal Z}_a) \ee where ${\cal Z}_a =
(z_a|\eta_a)$ are the super momentum-twistor variabes. This is
precisely the formula for computing on-shell diagrams (in
momentum-twistor space) as described in \cite{N4,LD1,N6}. Thus,
while the amplituhedron geometry and the associated form $\Omega$
are purely bosonic, we have extracted from them super-amplitudes
which are manifestly supersymmetric. Indeed, the connection to the
Grassmannian shows much more--the superamplitude obtained for each
cell is manifestly Yangian invariant \cite{N6}.

\section{Hiding Particles $\to$ Loop Positivity in $G_+(k,n;L)$}
The direct generalization of ``convex polygons" into the
Grassmannian $G(k,k+4)$ has given us the tree amplituhedron. We will
now ask a simple question: can we ``hide particles" in a natural
way? This will lead to an extended notion of positivity giving us
loop amplitudes.

It is trivial to imagine what we might mean by hiding a single
particle, but as we will see momentarily, the idea of hiding
particles is only natural if we hide {\it pairs} of {\it adjacent}
particles. To pick a concrete example, suppose we have some positive
matrix $C$ with columns we'll label $(A_1,B_1, 1, 2, \dots, m, A_2, B_2,
m+1, \dots n)$. We can always gauge-fix the $A_1,B_1$ and $A_2,B_2$ columns
so that the matrix takes the form

$$\bordermatrix{\text{}&A_1&B_1&1&2&\ldots&m&A_2&B_2&m+1&\ldots&n\cr
               &1&0&\ast&\ast&\ldots&\ast&0&0&\ast&\ldots & \ast\cr
               &0&1&\ast&\ast&\ldots&\ast&0&0&\ast&\ldots & \ast\cr
               &0&0&\ast&\ast&\ldots&\ast&1&0&\ast&\ldots & \ast\cr
               &0&0&\ast&\ast&\ldots&\ast&0&1&\ast&\ldots & \ast\cr
               &0&0&\ast&\ast&\ldots&\ast&0&0&\ast&\ldots & \ast\cr
               &\vdots&\vdots&\vdots&\vdots&\vdots&\vdots&\vdots&\vdots&\vdots&\vdots&\vdots\cr
               &0&0&\ast&\ast&\ldots&\ast&0&0&\ast&\ldots & \ast\cr}$$
We would now like to ``hide" the particles $A_1,B_1,A_2,B_2$. We do
this simply by chopping out the corresponding columns. The remaining
matrix can be grouped into the form \be \left(\begin{array}{ccc}&
D_{(1)} & \\ \hdashline  & D_{(2)} & \\ \hdashline & C & \end{array}
\right) \ee But the ``hidden" particles leave an echo in the
positivity properties of this matrix. The positivity of the minors
involving all of $(A_1,B_1,A_2,B_2)$, $(A_2,B_2)$ and $(A_1,B_1)$
individually, as well those not involving $A_1,B_1,A_2,B_2$ at all
enforce that the ordered maximal minors of the following matrices
\be \left(\begin{array}{ccc} & C & \end{array} \right), \left(
\begin{array}{ccc}& D_{(1)}& \\ \hdashline & C & \end{array} \right),
\left( \begin{array}{ccc} &D_{(2)}& \\ \hdashline & C  & \end{array}
\right), \left(\begin{array}{ccc} &  D_{(1)}&  \\ \hdashline &
D_{(2)}&
\\ \hdashline &C& \end{array} \right) \ee are all positive.

We can now see why particles are most naturally hidden in pairs. If
we had instead hidden single particles as $A_1, A_2, A_3, \dots$ in
separate columns, the remaining minors would be positive or negative
depending on the orderings of $A_1,A_2,A_3, \dots$, which is
additional structure over and above the cyclic ordering of the
external data. In order to avoid this arbitrariness, we should hide
particles in even numbers, with pairs the minimal case. In order to
ensure that only minors involving the pairs $(A_i B_i)$ are taken into
account,  we mod out by the $GL(2)$ action rotating the $(A_i,B_i)$ columns
into each other.

This ``hidden particle" picture has thus motivated an extended
notion of positivity associated with the Grassmannian. We are used
to considering a $k$-plane in $n$ dimensions $C$, with all ordered
minors positive. But we can also imagine a collection of $L$
2-planes $D_{(i)}$ in the $(n-k)$ dimensional complement of $C$,
schematically
$$
\includegraphics[scale=.65]{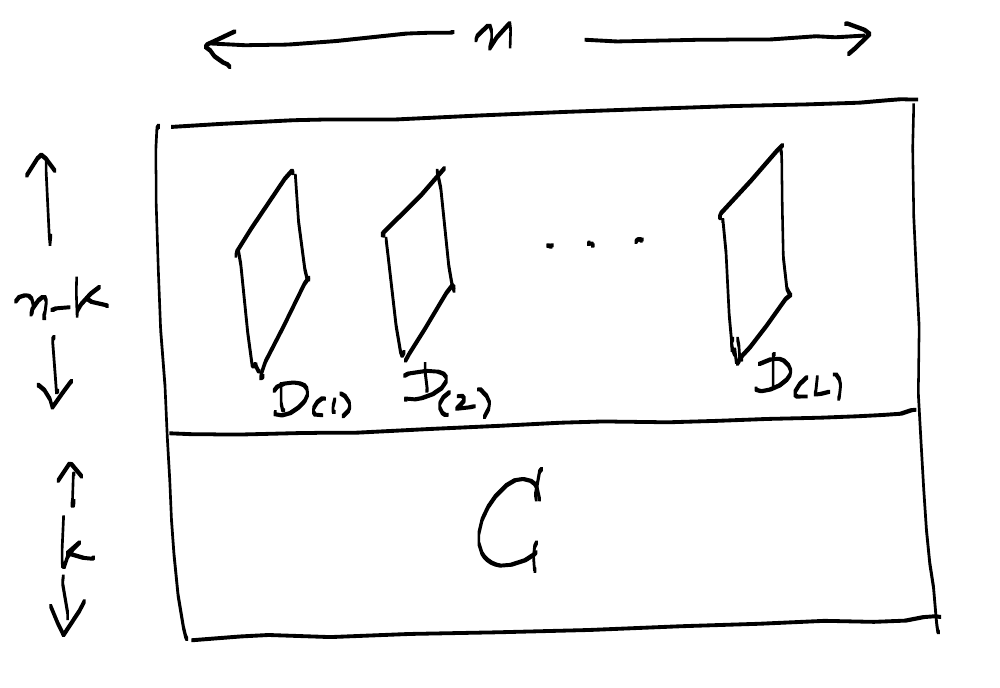}
$$
We call this space $G(k,n;L)$, and we will denote the collection of
$(D_{(i)}, C)$ as ${\cal C}$.  We can extend the notion of
positivity to $G(k,n;L)$ by demanding that not only the ordered
minors of $C$, but also of $C$ with any collection of the $D_{(i)}$,
are positive. (All minors must include the matrix $C$, since the
$D_{(i)}$ are defined to live in the complement of $C$). Note that
this notion is completely permutation invariant in the $D_{(i)}$.

Very interestingly, it turns out that while we motivated this notion
of positivity by hiding particles from an underlying positive
matrix, there are positive configurations of ${\cal C}$ that can not
be obtained by hiding particles from a positive matrix in this way.

Extending the map $Y = C . Z$ in the obvious way to include the
$D$'s leads us to define the full amplituhedron.

\section{The Amplituhedron ${\cal A}_{n,k,L}(Z)$}

We can now give the full definition of the amplituhedron ${\cal
A}_{n,k,L}(Z)$. First,  the external data for $n \geq k+4$
particles is given by the vectors $Z_a^I$ living in a $(4 + k)$
dimensional space; where $a= 1, \dots, n$ and $I = 1, \dots, 4 + k$.
The data is positive \be \langle Z_{a_1} \dots Z_{a_{4 + k}} \rangle
> 0 \quad {\rm for} \quad  a_1 < \dots < a_{4 + k} \ee The amplituhedron
lives in $G(k,k+4;L)$: the space of $k$ planes $Y$ in $(k+4)$
dimensions, together with $L$ 2-planes ${\cal L}_{(i)}$ in the 4
dimensional complement of $Y$, schematically
$$
\includegraphics[scale=.65]{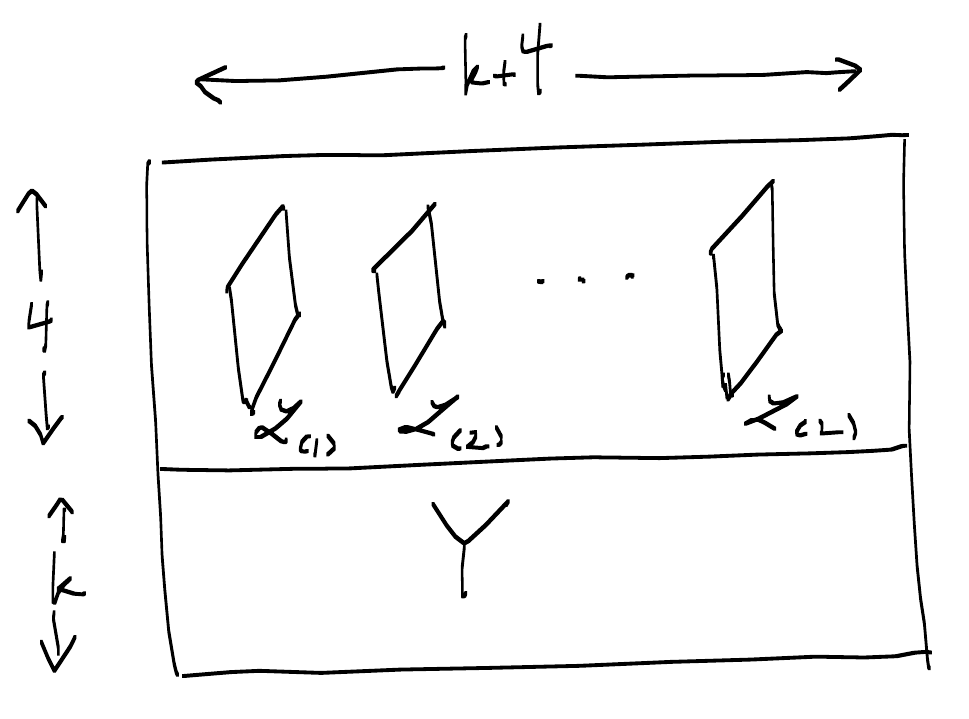}
$$
We will denote the collection of $({\cal L}_{(i)}, Y)$ as ${\cal
Y}$.

The amplituhedron ${\cal A}_{n,k,L}(Z)$ is the subspace of
$G(k,k+4;L)$ consisting of all ${\cal Y}$'s which are a positive
linear combination of the external data, \be {\cal Y} = {\cal C}
\cdot Z \ee More explicitly in components, the $k$-plane is
$Y_{\alpha}^I$, and the 2-planes are ${\cal L}_{\gamma (i)}^I$,
where $\gamma = 1,2$ and  $i = 1, \dots, L$ . The amplituhedron is
the space of all $Y, {\cal L}_{(i)}$ of the form \be Y_{\alpha}^I =
C_{\alpha a} Z_a^I, \, \, {\cal L}_{\gamma (i)}^{I} = D_{\gamma a
(i)} Z_a^I \ee where as in the previous section the $C_{\alpha a}$
specifies a $k$-plane in $n$-dimensions, and the $D_{\gamma a (i)}$
are $L$ 2-planes living in the $(n-k)$ dimensional complement of
$C$, with the positivity property that for any $0 \leq l \leq L$,
all the ordered $(k + 2l) \times (k+2l)$ minors of the $(k + 2
l) \times n$ matrix \be \left(\begin{array}{ccc} & D_{(i_1)} & \\
\hdashline & \vdots &  \\ \hdashline & D_{(i_l)} & \\ \hdashline & C
\end{array} \right) \ee are positive.

The notion of cells, cell decomposition and canonical form
can be extended to the full amplituhedron. A cell $\Gamma$ is
associated with a set of positive co-ordinates $\alpha^\Gamma =
(\alpha^\Gamma_1, \dots, \alpha^\Gamma_{4(k + L)})$, rational in the ${\cal C}$, such that for $\alpha$'s positive,
${\cal C}(\alpha) = (D_{(i)}(\alpha), C(\alpha))$ is in $G_+(k,n;L)$.  A
cell decomposition is a collection $T$ of non-intersecting cells
$\Gamma$ whose images under ${\cal Y} = {\cal C} \cdot Z$ cover the
entire amplituhedron. The rational form $\Omega_{n,k,L}({\cal Y}; Z)$ is
defined by having the property that

\begin{center}
$\Omega_{n,k,L}(Y;Z)$ has logarithmic singularities on all the
boundaries of ${\cal A}_{n,k,L}(Z)$
\end{center}

A concrete formula follows from a cell decomposition as \be
\Omega_{n,k,L}({\cal Y}; Z) = \sum_{\Gamma \subset T} \prod_{i =
1}^{4(k + L)} \frac{d \alpha_i^\Gamma}{\alpha_i^\Gamma} \ee Of course any cell
decomposition gives the same form $\Omega_{n,k,L}$.

\section{The Loop Amplitude Form}

We can extract the $4L$-form for the loop integrand from
$\Omega_{n,k,L}$ in the obvious way. The 2-planes ${\cal L}_{(i)}$,
being in the complement of $Y_0$, can be taken to be non-vanishing
in the first 4 entries ${\cal L}^I_{ (i)}  = ({\cal L}_{(i) 2 \times
4} | 0_{2 \times k})$. Each ${\cal L}_{\gamma (i)}$ gives us a line
$({\cal L}_{\gamma =1} {\cal L}_{\gamma = 2})_{(i)}$ (which we have
also been calling $(AB)_{(i)}$) in $\mathbb{P}^3$. These are the
momentum-twistor representation of the loop integration variables.
The analog of equation (\ref{super}) for the loop integrand form is
\be {\cal M}_{n,k}(z_a,\eta_a; {\cal L}_{(\gamma (i)}) = \int d^{4}
\phi_1 \dots d^4 \phi_k \int \Omega_{n,k,L}(Y,{\cal L}_{\gamma
(i)};Z) \delta^{4k}(Y;Y_0) \ee Any form on $G(k,k+4k;L)$ can be
written as
\begin{equation}
\Omega =  \langle Yd^4 Y_1 \rangle \dots \langle Yd^4 Y_k \rangle
\prod_{i=1}^L \langle Y {\cal L}_{1(i)} {\cal L}_{2(i)} d^2 {\cal
L}_{1(i)} \rangle \langle Y {\cal L}_{1(i)} {\cal L}_{2(i)} d^2
{\cal L}_{2(i)} \rangle \times \omega_{n,k,L}(Y,{\cal L}_{(i)})(Z)
\end{equation}
where we denoted $Y=Y_1\dots Y_k$. So we have for the integrand of
the all-loop amplitude
\begin{align}
{\cal M}_{n,k}(z_a,\eta_a,{\cal L}_{\gamma(i)}) &= \int d^4 \phi_1
\dots d^4 \phi_k  \prod_{i=1}^L \langle {\cal L}_{1(i)} {\cal
L}_{2(i)} d^2 {\cal L}_{1(i)} \rangle \langle  {\cal L}_{1(i)} {\cal
L}_{2(i)} d^2 {\cal L}_{2(i)} \rangle \omega_{n,k}(Y_0,{\cal
L}_{\gamma(i)};Z_a)
\end{align}
Already the simplest case $k=0$ of the amplituhedron is interesting
at loop level. At 1-loop, we have a 2-plane in 4 dimensions $AB$,
and  the $D$ matrix is just restricted to be in $G_+(2,n)$. It is
easy to see that the 4 dimensional cells of $G_+(2,n)$ are labeled
by a pair of triples $[a,b,c;x,y,z]$, where the top row of the
matrix is non-zero in the columns $(a,b,c)$ and the bottom in
columns $(x,y,z)$. A simple collection of these \be \sum _{i<j}
[1\,i\,i\pl1;\,1\,j\,j\pl1] \ee beautifully covers the amplituhedron
in this case. The map into $G(2,4)$ for each cell is \be A = Z_1 +
\alpha_i Z_i + \alpha_{i+1} Z_{i+1}, \, B  = -Z_1 + \alpha_j Z_j +
\alpha_{j+1} Z_{j+1} \ee and so the form associated with the cell is
\begin{align} &\frac{d
\alpha_i}{\alpha_i} \frac{d\alpha_{i+1}}{\alpha_{i+1}}
\frac{\alpha_j}{\alpha_j} \frac{d \alpha_{j+1}}{\alpha_{j+1}} =
\frac{\langle AB d^2 A \rangle \langle AB d^2 B \rangle \langle AB
(1\,i\,i\pl1) \cap (1\,j\,j\pl1) \rangle^2}{\langle AB\,1\,i \rangle
\langle AB\,1\,i\pl1 \rangle \langle AB\,i\,i\pl1 \rangle \langle
AB\,1\,j\rangle \langle AB\,1\,j\pl1 \rangle \langle AB\,j\,j\pl1
\rangle}
\end{align}
The form $\Omega$ gives exactly the ``Kermit" expansion for the MHV
integrand given in \cite{N5}, now obtained without any reference to tree
amplitudes, forward limits or recursion relations.

In this simple case, direct triangulation of the space is straightforward. But we could also have worked backwards, starting with the BCFW formula, and  recognizing how each term in the ``Kermit" expansion is associated with positive co-ordinates for some cell of the amplituhedron. We could then observe that, remarkably,  these cells are non-overlapping, and together cover the full amplituhedron.

In order to illustrate more of the structure of the loop
amplituhedron, including the interplay between the $``C"$ and $``D"$
matrices, let us consider the 1-loop $k=1$ amplitude for $n=6$.
There are 16 terms in the BCFW recursion, which can all be mapped
back to their $Y, AB$ space form, and in turn associated with
positive co-ordinates in the amplituhedron. For instance, one of
BCFW terms is

$$
\frac{\la YAB13\ra\la YAB(561)\cap(2345)\ra^4\la
YAB(123)\cap(Y456)\ra^2}{\begin{array}{c} \la Y2345\ra\la
YAB(561)\cap(Y345)\ra\la YAB(561)\cap(Y234)\ra\la
YAB(561)\cap(Y235)\ra\la YAB56\ra \\ \la
YAB(561)\cap(Y45(23)\cap(YAB1))\ra\la YAB12\ra\la YAB23\ra\la
YAB13\ra\la YAB15\ra\la YAB16\ra\end{array}}
$$
While it may not be immediately apparently, this is nothing but the
``dlog" canonical form associated with the following positive
co-ordinates for the $(D,C)$ matrix
$$
\left(\begin{array}{ccc} & D & \\ \hdashline & C & \end{array}
\right) = \left(
  \begin{array}{cccccc}
1 & x & y & 0 & 0 & 0 \\
 -w & 0 & 0 & 0 & -1 & -z \\ \hdashline
 w & xt_1 & t_2+t_1y & t_3 & 1+t_4 & z \\
  \end{array}
\right)
$$
This exercise can be repeated with all $16$ BCFW terms. The
corresponding $(D,C)$ matrices are
$$
\hspace{-0.5cm}\includegraphics[scale=1]{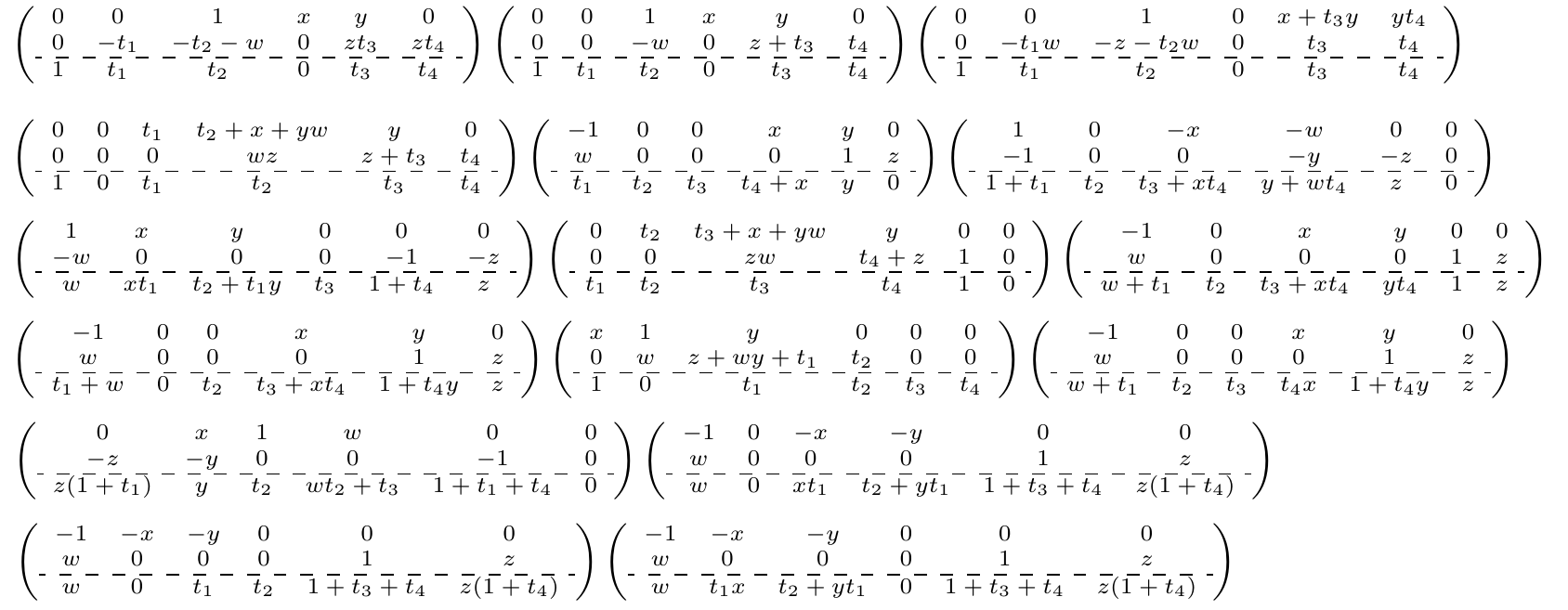}
$$
One can easily check that for all variables positive, the bottom row
of these matrices is positive, and all the ordered  $3 \times 3$
minors are also positive. For any cell, we can range over all the
positive variables, which under the ${\cal Y} = {\cal C} \cdot Z$
gives an image of the cell in $(Y,AB)$ space. Remarkably, we find
that these cells are non-overlapping, and cover the entire space.
This can be checked directly in a simple way. We begin by fixing
positive external data $(Z_1, \cdots, Z_6)$. We then choose any
positive matrix ${\cal C}$ at random, which gives an associated
point ${\cal Y}$ inside the amplituhedron. We can ask whether or not
this point is contained in one of the cells, by seeing whether
${\cal Y}$ can be reproduced with positive values for all eight
variables of the cell. Doing this we find that every point in the
amplituhedron is contained in just one of these cells (except of
course for points on the common boundaries of different cells). The
cells taken together therefore give a cellulation of the
amplituhedron.

Note that the form shown above, associated with a BCFW term, has some
physical poles (like $\langle Y AB 12 \rangle$), but also many
unphysical poles. The unphysical poles are associated with
boundaries of the cell that are ``inside" the amplituhedron, and not
boundaries of the amplituhedron themselves. These boundaries are
spurious, and so are the corresponding poles, which cancel in the
sum over all BCFW terms.

We have checked in many other examples, for higher $k$ and also at
higher loops, that $(a)$ BCFW terms can be expressed as canonical
forms associated with cells of the amplituhedron and $(b)$ these
collection of cells do cover the amplituhedron.

It is satisfying to have a definition of the loop amplituhedron that
lives directly in the space relevant for loop amplitudes. This is in
contrast with the approach to computing the loop integrand using
recursion relations, which ultimately traces back to higher $k$ and
$n$ tree amplitudes. Consider the simple case of the 2-loop
4-particle amplitude. We are after a form in the space of two
2-planes $(AB)_1, (AB)_2$ in four dimensions. The BCFW approach begins with
the $k=2, n=8$ tree amplitudes, and arrives at the form we are
interested in after taking two ``forward limits". But the
amplituhedron lives directly in the $(AB)_1, (AB)_2$ space, and we can find
a cell decomposition for it directly, yielding the form without
having to refer to any tree amplitudes.

We have understood how to directly ``cellulate" the amplituhedron in
a number of other examples, and strongly suspect that there will be
a general understanding for how to do this. The BCFW
decomposition of tree amplitudes seems to be associated with
particularly nice, canonical cellulations of the tree
amplituhedron. Loop level BCFW also gives a cell decomposition. The ``direct" cellulations we have
found in many cases are however simpler, without an obvious connection to the BCFW expansion.

\section{Locality and Unitarity from Positivity}

Locality and unitarity are encoded in the positive geometry of the
amplituhedron in a beautiful way.  As is well-known, locality and
unitarity are directly reflected in the singularity structure of the
integrand for scattering amplitudes. In momentum-twistor language,
the only allowed singularities at tree-level should occur when
$\langle Z_i Z_{i+1} Z_j Z_{j+1} \rangle \to 0$; in the loop-level
integrand, we can also have poles of the form $\langle AB\,i\,i\pl1
\rangle \to 0$, and $\langle AB_{(i)} AB_{(j)} \rangle \to 0$.
Unitarity is reflected in what happens as  poles are approached,
schematically we have \cite{N6}
$$
\includegraphics[scale=.8]{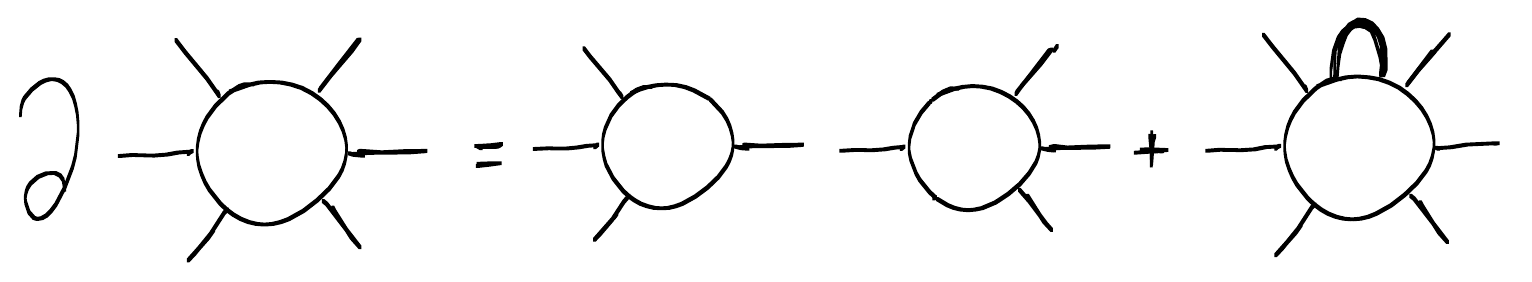}
$$

Given the connection between the form $\Omega_{n,k,L}$ and the
amplitude, it is obvious that the first (co-dimension one) poles of
the amplitude are associated with the co-dimension one ``faces" of
the amplituhedron. For trees, we have already seen that, remarkably,
positivity forces these faces to be precisely where $\langle Y_1
\dots Y_k Z_i Z_{i+1} Z_j Z_{j+1} \rangle \to 0$, exactly as needed
for locality. The analog statement for the full loop amplituhedron
also obviously includes $\langle Y_1 \cdots Y_k AB\,i\,i\pl1 \rangle
\to 0$.

The factorization properties of the amplitude also follow directly
as a consequence of positivity. For instance, let us consider the
boundary of the tree amplituhedron where the $k$ plane $(Y_1 \dots
Y_k)$ is on the plane $(Z_i Z_{i+1} Z_j Z_{j+1})$. We can e.g.
assume that $Y_1$ is a linear combination of $(Z_i, Z_{i+1}, Z_j,
Z_{j+1})$, and thus that the top row of the $C$ matrix is only
non-zero in these columns. But then, positivity remarkably forces
the $C$ matrix to  ``factorize" in the form
$$
\includegraphics[scale=1]{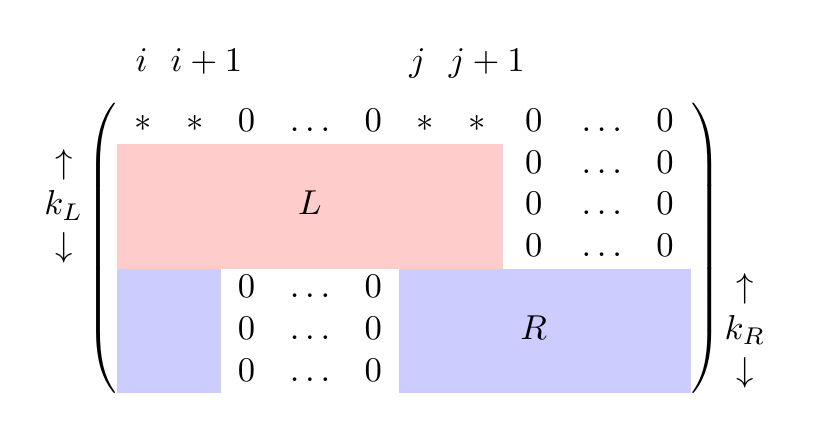}
$$
for all possible $k_L, k_R$ such that $k_L + k_R = k - 1$. This
factorized form of the $C$ matrix in turn implies that on this
boundary, the amplituhedron does ``split" into lower-dimensional
amplituhedra in exactly the way required for the factorization of
the amplitude.

We can similarly understand the single-cut of the loop integrand.
Consider for concreteness the simplest case of the $n$ particle
one-loop MHV amplitude. On the boundary where $\langle AB\,n1
\rangle \to 0$, the $D$ matrix has the form

$$\bordermatrix{\text{}&1&2& \dots &n\cr
               &1&0& \dots & -x_n\cr
               &y_1&y_2&\dots&y_n \cr}$$

The connection of this $D$ matrix to the forward limit
\cite{CaronHuot:2010zt} of the NMHV tree amplitude is simple. In the
language of \cite{N5}, the forward limit in momentum-twistor space
is represented as
$$
\includegraphics[scale=.55]{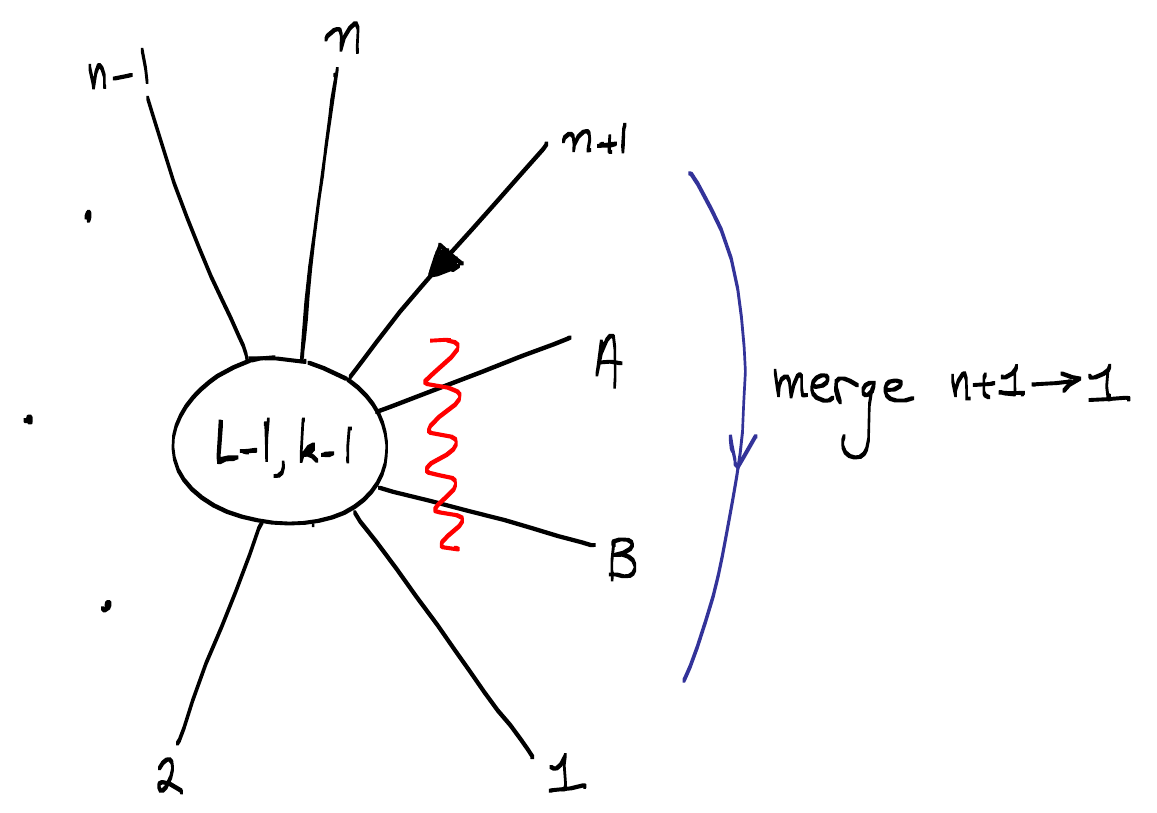}
$$
we start with the tree NMHV amplitude, associated with the positive
$1 \times n$ matrix \be (y_A \, y_B \, y_1 \, y_2 \, \dots \, y_n)
\ee and first we ``add" particle $n+1$ between $n$ and $A$, which
adds three degrees of freedom $x_n, x_A, \alpha$
$$\bordermatrix{\text{}&A&B&1&2& \dots &n & n+1\cr
               &x_A& \alpha x_A & 0 & 0 & \dots & -x_n & -1\cr
               &y_A&y_B + \alpha y_A& y_1 & y_2 & \dots&y_n & 0\cr}$$
and we finally ``merge" $n+1,1$, which means shifting column 1 as
$c_1 \to c_ 1 - c_{n+1}$ and removing column $(n+1)$. This gives us
the matrix

$$\bordermatrix{\text{}&A&B&1&2& \dots &n\cr
               &x_A& \alpha x_A & 1 & 0 & \dots & -x_n \cr
               &y_A&y_B + \alpha y_A& y_1 & y_2 & \dots&y_n\cr}$$
note that the the $A,B$ columns have precisely four degrees of
freedom $x_A,\alpha, y_A, y_B$ which we can remove by $GL(2)$ acting
on the $A,B$ columns. Chopping off $A,B$ we are then left precisely
with the $D$ matrix on the single cut. This shows that the single cut of the loop integrand is the forward limit
of the tree amplitude, exactly as required by unitarity.

\section{Four Particles at All Loops}

Let us briefly describe the  simplest example illustrating the novelties of positivity at
loop level, for four-particle scattering at $L$
loops. We can parametrize each $D_{(i)}$ as \be D_{(i)} =
\left(\begin{array}{cccc} 1 & x_i & 0 & -w_i \\ 0 & y_i & 1 & z_i
\end{array} \right) \ee In this simple case the positivity
constraints are just that all the $2 \times 2$ minors of $D_{(i)}$
and the $4 \times 4$ minors \be {\rm det} \left(\begin{array}{ccc} &
D_{(i)} & \\ \hdashline & D_{(j)} & \end{array} \right) \ee are
positive. This translates to \be x_i, y_i, z_i, w_i > 0, \quad (x_i
- x_j)(z_i - z_j) + (y_i - y_j)(w_i - w_j) < 0 \ee We can rephrase
this problem in a  simple, purely geometrical way by  defining
two dimensional vectors $\vec{a}_i = (x_i,y_i), \vec{b}_i = (z_i,
w_i)$. The points are in the upper quadrant of the plane. The mutual
positivity condition is just $(\vec{a}_i - \vec{a}_j) \cdot
(\vec{b}_i - \vec{b}_j) < 0$. Geometrically this just means that the
$\vec{a}, \vec{b}$ must be arranged so that for every pair $i,j$,
the line directed from $\vec{a}_i \to \vec{a}_j$ is pointed in the
opposite direction as the one directed from $\vec{b}_i \to
\vec{b}_j$. An example of an allowed configuration of such points
for $L=3$ is
$$
\includegraphics[scale=.6]{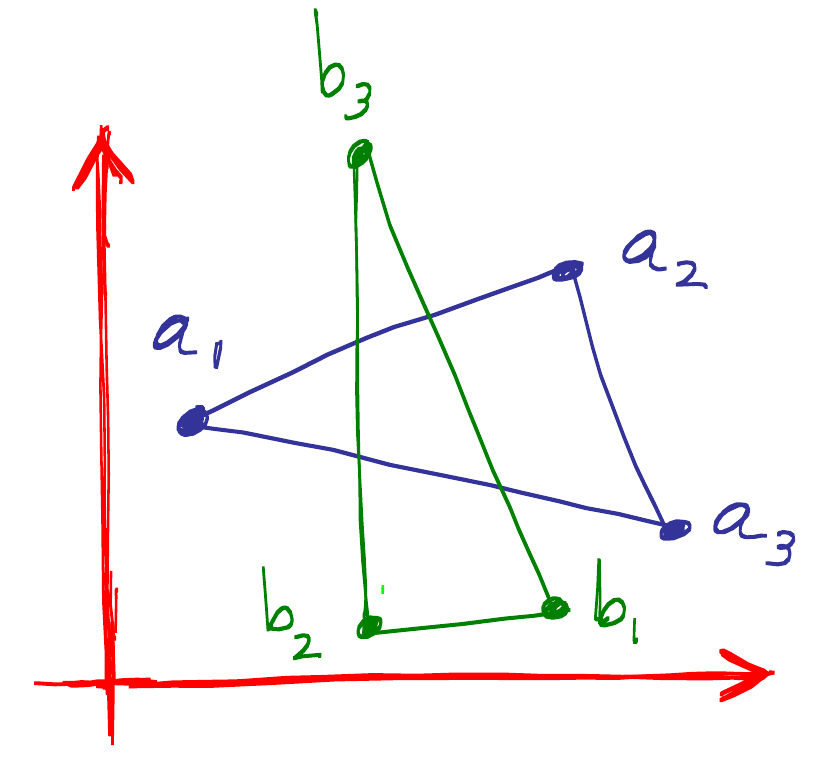}
$$
Finding a cell decomposition of this $4L$ dimensional space directly
gives us the integrand for the four-particle amplitude at $L$-loops.

Now, we know that the final form can be expressed as a sum over
local, planar diagrams. This makes it all the more remarkable that
no-where in the definition of our geometry problem do we
reference to diagrams of any sort, planar or not!
Nonetheless, this property is one of many that emerges from
positivity.

As we will describe at greater length in \cite{Into}, it is easy to
find a cell decomposition for the full space ``manually" at low-loop
orders. We suspect there is a more systematic approach to
understanding the geometry that might crack the problem at all loop
order. As an interesting  warmup to the full problem, we can
investigate lower-dimensional ``faces" of the four-particle amplituhedron. Cellulations
of these faces corresponds to computing certain
cuts of the integrand, at all loop orders. We will discuss many of
these faces and cuts systematically in \cite{Into}. Here we will
content ourselves by presenting some especially simple but not
completely trivial examples.

Let us start by considering an extremely simple boundary of the space,
where all the $w_i \to 0$. This corresponds to having all the lines
intersect $(Z_1 Z_2)$. The positivity conditions then simply become
\be (x_i - x_j)(z_i - z_j) < 0 \ee  which is trivial to
triangulate. Whatever configuration of $x$'s we have are ordered in
some way, say $x_1 < \dots < x_L$.  Then we must have $z_1 > \dots
> z_L$. The $y_i$ just have to be positive. The associated form is
then trivially (we omit the measure $\prod_i dx_i dz_i d y_i$): \be
\frac{1}{y_1} \dots \frac{1}{y_L}  \frac{1}{x_1} \frac{1}{x_2 - x_1}
\dots \frac{1}{x_L - x_{L-1}} \frac{1}{z_L} \frac{1}{z_{L-1} - z_L}
\dots \frac{1}{z_1 - z_2} + {\rm perm.} \ee Now, this cut is
particularly simple to understand from the point of view of the
familiar ``local" expansions of the integrand--there is only  only
local diagram that can possibly contribute to this cut: the
``ladder" diagram. The corresponding cut is precisely what we
have above from positivity.
$$
\includegraphics[scale=.8]{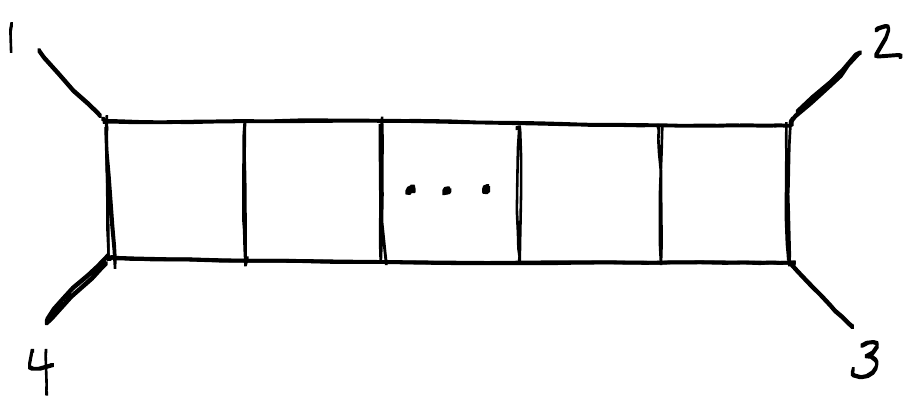}
$$

We can continue along these lines to explore  faces of the
amplituhedron which determine cuts to all loop orders that are
difficult (if not impossible) to derive in any other way. For
instance, suppose that some of the lines  intersect $(Z_1
Z_2)$, so that the $w_i \to 0$ for $i = 1, \dots, L_1$ and others
intersects $(Z_3 Z_4)$, so that $y_I \to 0$ for $I =L_1+1, \dots,
L$. To pick a concrete interesting example, let choose $L-2$ lines
to intersect $(12)$ and 2 lines to intersect $(34)$. We can further
specialize the geometry and take more cuts by making the $L$'th line
pass through the point 3 -- this corresponds to sending $z_{L} \to
0$. Let us also take the $(L-1)$'st line to pass through the point 4
-- this corresponds to sending $z_{L-1}, w_{L-1} \to \infty$ with
$w_{L-1}/z_{L-1} \equiv W_{L-1}$ fixed.

We can again label the $x_i; x_I$ so they are in increasing order;
then the positivity conditions become \be x_1 < \dots < x_{L-2}, z_1
> \dots > z_{L-2}; \, x_{L-1} < x_{L} \ee and \be W_{L-1} y_i >
(x_{L-1} - x_i), \, \, w_L y_i > z_i (x_i - x_L) \ee This space is
also trivial to triangulate, but the corresponding form is more
interesting. The ordering for the $z$'s is associated with the form
$$
\frac{1}{z_{L\mi2}(z_{L\mi3}-z_{L\mi2})(z_{L\mi4}-z_{L\mi3})\dots
(z_1-z_2)}
$$
The interesting part of the space involves $x_i,y_i$. Note that if $x_i
< x_{L-1}$, the second inequality on $y_i$ is trivially satisfied
for positive $y_i$, and the only constraint on $y_i$ is just $y_i >
(x_{L-1} - x_i)/W_{L-1}$. If $x_{L-1} < x_i < x_{L}$, then both
inequalities are satisfied and we just have $y_i > 0$. Finally if
$x_i > x_L$, the first inequality is trivially satisfied and we just
have $y_i > z_i (x_i - x_L)/w_L$. Thus, given any ordering for all
the $x's$, there is an associated set of inequalities on the $y$'s,
and the corresponding form in $x,y$ space is trivially obtained. For
instance, consider the case $L=5$, and an ordering for the $x$'s
where $x_1 < x_4 < x_2 < x_5 < x_3$. The corresponding form in
$(x,y)$ space is just
\begin{align}
\frac{1}{x_1 (x_4 - x_1) (x_2 - x_4) (x_5 - x_2)(x_3 - x_5)}
\frac{1}{y_1 - (x_4 - x_1)/W_4} \frac{1}{y_2} \frac{1}{y_3 - z_3(x_3
- x_5)/w_5}
\end{align}
By summing over all the possible orderings $x$'s, we get the final
form. For general $L$, we can simply express the result (again
omitting the measure) as a sum over permutations $\sigma$:

\begin{align}
&\prod_{l=1}^{L-2}\frac{1}{(z_l -
z_{l+1})}\quad\times\hspace{-0.7cm}\sum_{\sigma; \sigma_1 < \dots <
\sigma_{L-2}; \sigma_{L-1} <
\sigma_L} \frac{1}{w_L W_{L-1}} \prod_{l=1}^{L} \frac{1}{(x_{\sigma^{-1}_l} - x_{\sigma^{-1}_{l-1}})}\\
&\hspace{4cm}\times\prod_{i=1}^{L-2} \left\{ \begin{array}{cc}
(y_i - (x_{L-1} - x_i)/W_{L-1})^{-1} &  \sigma_i < \sigma_{L-1} \\
y_i^{-1} &  \sigma_{L-1} < \sigma_i < \sigma_{L} \\ (y_i - (x_i -
x_L) z_i/w_L)^{-1} &  \sigma_L < \sigma_i  \end{array}\right\}
\nonumber
\end{align}
where we define for convenience $z_{L-1} = x_{\sigma^{-1}_0} = 0$.

This gives us non-trivial all-loop order information about
the four-particle integrand. The expression has a feature familiar
from BCFW recursion relation expressions for tree and loop level
amplitudes. Each term has certain ``spurious" poles, which cancel in
the sum. This result can be checked against the cuts of the
corresponding amplitudes that are available in ``local form". The
diagrams that contribute are of the type
$$
\includegraphics[scale=.7]{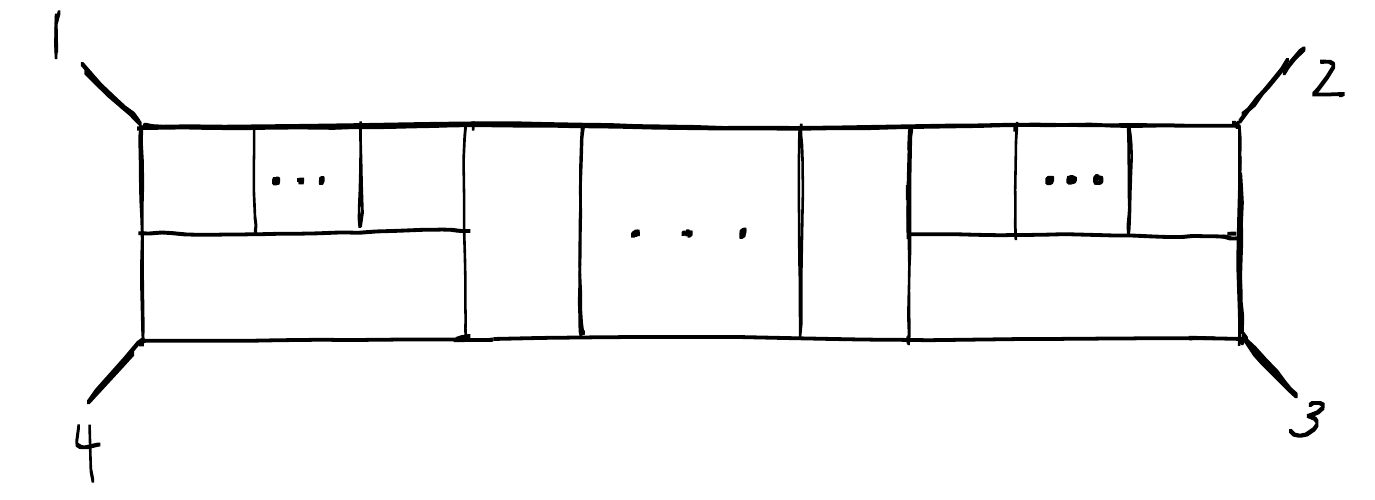}
$$
but now there are non-trivial numerator factors that don't trivially
follow from the structure of propagators. The full integrand is
available through to seven loops in the literature
\cite{Bern:2005iz,Bern:2006ew,Bern:2007ct,Bourjaily:2011hi,Eden:2012tu}.
The inspection of the available  local expansions on this cut does
not indicate an obvious all-loop generalization, nor does it betray
any hint that that the final result can be expressed in the one-line
form given above. For instance just at 5 loops, the local form of
the cut is given as a sum over diagrams,
$$
\includegraphics[scale=.8]{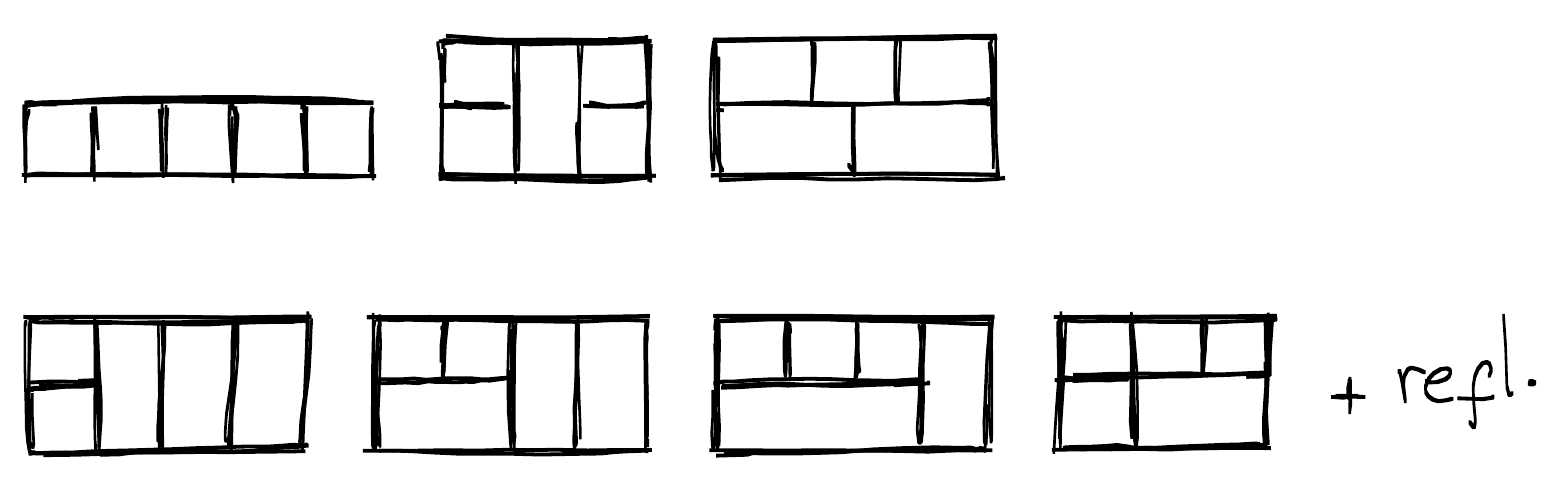}
$$
with intricate numerator factors. If all terms are combined with a
common denominator of all physical propagators, the numerator has
347 terms. Needless to say, the complicated expression obtained in
this way perfectly matches the amplituhedron computation of the cut.

\section{Master Amplituhedron}
We have defined the amplituhedron ${\cal A}_{n,k,L}$ separately for
every $n,k$ and loop order $L$. However, a trivial feature of the
geometry is that ${\cal A}_{n,k,L}$ is contained in the ``faces" of
${\cal A}_{n^\prime, k^\prime, L^\prime}$, for $n^\prime >n,
k^\prime > k, L^\prime > L$. The objects needed to compute
scattering amplitudes for any number of particles to all loop orders
are thus contained in a ``master amplituhedron" with $n, k , L \to
\infty$.

In this vein it may also be worth considering natural
mathematical generalizations of the amplituhedron. We have already seen
that the generalized tree amplituhedron ${\cal A}_{n,k,m}$ lives in $G(k,k+m)$ and can be
defined for any even $m$. It is obvious that the amplituhedron with
$m=4$, of relevance to physics, is contained amongst the faces of
the object defined for higher $m$.

If we consider general even $m$, we can also generalize the notion of
``hiding particles" in an obvious way: adjacent particles can be hidden
in even numbers. This leads us to a bigger space in which
to embed the generalized loop amplituhedron. Instead of just considering
$G(k,k+4;L)$ of ($k-$ planes) $Y$ together with $L$ ($2-$planes) in
$m=4$ dimensional complement of $Y$, we can consider a space
$G(k,k+m;L_2,L_4, \dots, L_{m-2})$, of $k$-planes $Y$ in $(k+m)$
dimensions, together with $L_2$ (2-planes), $L_4$ (4-planes), $\dots
L_{m-2}$ ($(m-2)$-planes) in the $m$ dimensional complement of $Y$,
schematically:
$$
\includegraphics[scale=.6]{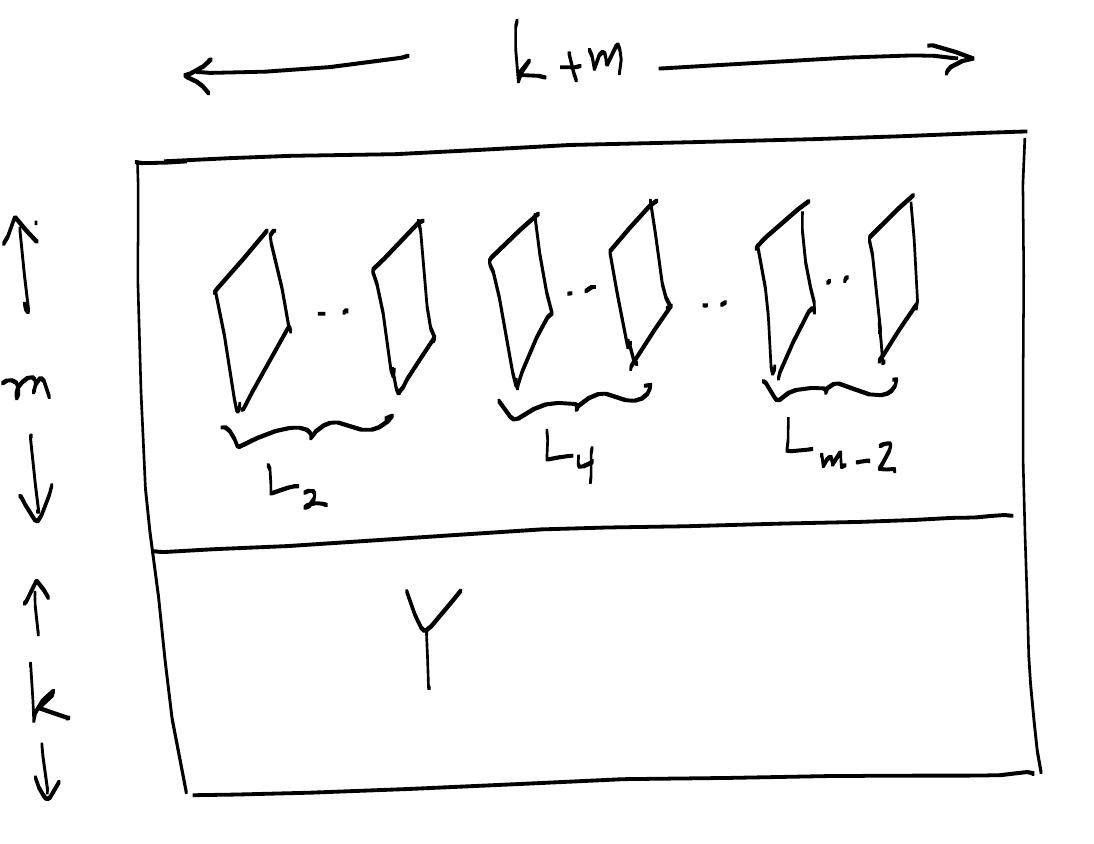}
$$
Once again we have ${\cal Y} = {\cal C} \cdot Z$, with the obvious
extension of the loop positivity conditions on ${\cal C}$ to
$G(k,n;L_2, L_4, \dots, L_{m-2})$. We can call this space ${\cal
A}_{n,k;m,L_2, \dots, L_{m-2}}(Z)$. The $m=4$ amplituhedron is again
just a particular face of this object. It would be interesting to
see whether this larger space has any interesting role to play in
understanding the $m=4$ geometry relevant to physics.

\newpage

\section{Outlook}

This paper has concerned itself with perturbative  scattering
amplitudes in gauge theories. However the deeper motivations for
studying this physics, articulated in \cite{N1,N2} have to do with
some fundamental challenges of quantum gravity. We have long known
that quantum mechanics and gravity together make it impossible to
have local observables. Quantum mechanics forces us to divide the
world in two pieces--an infinite measuring apparatus and a finite
system being observed. However for any observations made in a finite
region of space-time, gravity makes it impossible to make the
apparatus arbitarily large, since it also becomes heavier, and
collapses the observation region into a black hole. In some cases
like asymptotically AdS or flat spaces, we still have precise
quantum mechanical observables, that can be measured by infinitely
large apparatuses pushed to the boundaries of space-time: boundary
correlators for AdS space and the S-matrix for flat space. The fact
that no precise observables can be associated with the inside of the
space-time strongly suggests that there should be a way of computing
these boundary observables without any reference to the interior
space-time at all. For asymptotically AdS spaces, gauge-gravity
duality \cite{Maldacena:1997re} gives us a wonderful description of
the boundary correlators of this kind, and gives a first working
example of emergent space and gravity. However, this duality is
still an equivalence between ordinary physical systems described in
standard physical language, with time running from infinite past to
infinite future. This makes the duality inapplicable to our universe
for cosmological questions. Heading back to the early universe,  an
understanding of emergent time is likely necessary to make sense of
the big-bang singularity. More disturbingly, even at late times, due
to the accelerated expansion of our universe,  we only have access
to a finite number of degrees of freedom, and thus the division of
the world into ``infinite" and ``finite" systems, required by
quantum mechanics to talk about precise observables, seems to be
impossible \cite{Witten:2001kn}. This perhaps indicates the need for
 an extension of quantum mechanics to deal with subtle cosmological questions.

Understanding emergent space-time or possible cosmological
extensions of quantum mechanics  will obviously be a tall order. The
most obvious avenue for progress is directly attacking the
quantum-gravitational questions where the relevant issues must be
confronted. But there is another strategy that takes some
inspiration from the similarly radical  step taken in the transition
from classical to quantum mechanics, where classical determinism was
lost. There is a powerful clue to the coming of quantum mechanics
hidden in the structure of classical mechanics itself. While
Newton's laws are manifestly deterministic, there is a completely
different formulation of classical mechanics--in terms of the
principle of least action--which is not manifestly deterministic.
The existence of these very different starting points leading to the
same physics was somewhat mysterious to classical physicists, but
today we know why the least action formulation exists: the world is
quantum-mechanical and not deterministic, and for this reason, the
classical limit of quantum mechanics can't immediately land on
Newton's laws, but must match to some formulation of classical
physics where determinism is not a central but derived notion. The
least action principle formulation is thus much closer to quantum
mechanics than Newton's laws, and gives a better jumping off point
for making the transition to quantum mechanics as a natural
deformation, via the path integral.

We may be in a similar situation today. If there is a more
fundamental description of physics where space-time and perhaps even
the usual formulation of quantum mechanics don't appear, then even
in the limit where non-perturbative gravitational effects can be
neglected and the physics reduces to perfectly local and unitary
quantum field theory, this description is unlikely to directly
reproduce the usual formulation of field theory, but must rather
match on to some new formulation of the physics where locality and
unitarity are derived notions. Finding such reformulations of
standard physics might then better prepare us for the transition to
the deeper underlying theory.

In this paper, we have taken a baby first step in this direction,
along the lines of the  program put forward in \cite{N1,N2} and
pursued in \cite{N4,N5,N6}. We have given a formulation for planar
${\cal N} = 4$ SYM scattering amplitudes with no reference to
space-time or Hilbert space, no Hamiltonians, Lagrangians or gauge
redundancies, no path integrals or Feynman diagrams, no mention of
``cuts", ``factorization channels", or recursion relations. We have
instead presented  a new geometric question, to which the scattering
amplitudes are the answer. It is remarkable that such a  simple
picture, merely moving from ``triangles" to ``polygons", suitably
generalized to the Grassmannian, and with an extended notion of
positivity reflecting ``hiding" particles, leads us to the
amplituhedron ${\cal A}_{n,kL}$, whose ``volume" gives us the
scattering amplitudes for a non-trivial interacting quantum field
theory in four dimensions. It is also fascinating that while in  the
usual formulation of field theory, locality and unitarity are in
tension with each other, necessitating the introduction of the
familiar redundancies to accommodate both, in the new picture they
emerge together from positive geometry.

A great deal remains to be done both to establish and more fully
understand our conjecture. The usual  positive Grassmannian has a
very rich cell structure. The task of understanding all possible
ways to make ordered $k \times k$ minors of a $k \times n$ matrix
positive seems daunting at first, but the key is to realize that the
``big" Grassmannian can be obtained by gluing together
(``amalgamating" \cite{FG}) ``little" $G(1,3)$'s and $G(2,3)$'s,
building up larger positive matrices from smaller ones \cite{N6}.
Remarkably, this extremely natural mathematical operation translates
directly to the physical picture of building on-shell diagrams from
gluing together elementary three-particle amplitudes.  This story of
\cite{N6} is most naturally formulated in the original twistor space
or momentum space, while the amplituhedron picture  is formulated in
momentum-twistor space. At tree-level, there is a direct  connection
between the cells of $G(k,n)$ that cellulate the amplituhedron, and
those of $G(k+2,n)$, which give the corresponding on-shell diagram
interpretation of the cell \cite{N6}.  In this way, the natural
operation of decomposing the amplituhedron into pieces is ultimately
turned into a vivid on-shell scattering picture in the original
space-time.  Moving to loops, we don't have an analogous
understanding of all possible cells of the extended positive space
$G_+(k,n;L)$--we don't yet know how to systematically find positive
co-ordinates, how to think about boundaries and so on, though
certainly the on-shell representation of the loop integrand as
``non-reduced" diagrams  in $G(k+2,n)$ \cite{N6} gives hope that the
necessary understanding can be reached. Having control of  the cells
and positive co-ordinates for  $G_+(k,n;L)$ will very likely be
necessary to properly understand the cellulation ${\cal A}_{n,k;L}$.
It would also clearly be very illuminating to find an analog of the
amplituhedron, built around  positive external data in the original
twistor variables.This might also shed light on the connection
between these ideas and Witten's twistor-string theory
\cite{Witten:2003nn,Roiban:2004ix}, along the lines of
\cite{Dolan:2009wf, Nandan:2009cc, ArkaniHamed:2009dg,
Bourjaily:2010kw}.

While cell decompositions of the amplituhedron are geometrically
interesting in their own right, from the point of view of physics,
we need them only as a stepping-stone to determining the form
$\Omega_{n,k,L}$. This form was motivated by the idea of the area of
a (dual) polygon. For polygons, we have another definition of
``area", as an integral, and this gives us a completely invariant
definition for $\Omega$  free of the need for any triangulation. We
do not yet have an analog of the notion of ``dual amplituhedron",
and also  no integral representation for $\Omega_{n,k,L}$. However
in  \cite{Threeviews}, we will give strong circumstantial evidence
that such such an expression should exist. On a related note, while
we have a simple geometric picture for the loop integrand at any
fixed loop order, we still don't have a non-perturbative question to
which  the full amplitude (rather than just the fixed-order loop
integrand) is the answer.

Note that the form $\Omega_{n,k,L}$ is given directly by
construction as a sum of ``dlog" pieces. This is a highly
non-trivial property of the integrand, made manifest (albeit less
directly) in the on-shell diagram representation of the amplitude
\cite{N6} (see also \cite{Lipstein:2012vs, Lipstein:2013xra}).
 Optimistically, the great simplicity of this form should
allow a new picture for carrying out the integrations and arriving
at the final amplitudes. The crucial role that positive external
data played in our story suggests that this positive structure must
be reflected in the final amplitude in an important way. The
striking appearance of ``cluster variables" for external data in
\cite{posextamp} is an example of this.

We also hope that with a complete geometric picture for the
integrand of the amplitude in hand, we are now positioned to make direct
contact with the explosion of progress in using ideas from
integrability to determine the amplitude directly
\cite{CaronHuot:2011kk,Basso:2013vsa,Basso:2013aha, Dixon:2013eka}. A particularly promising place
to start forging this connection is with the four-particle amplitude
at all loop orders. As we noted, the positive geometry problem in
this case is especially simple, while the coefficient of the log$^2$
infrared divergence of the (log of the)  amplitude gives the cusp
anomalous dimension, famously determined using integrability
techniques in \cite{BS, Beisert:2006ez,Eden:2006rx}. Another natural question is how the introduction of the spectral
parameter in on-shell diagrams given in \cite{Ferro:2013dga,Ferro:2012xw} can be realized at the level of the amplituhedron.

On-shell diagrams in ${\cal N} = 4$ SYM and the positive
Grassmannian have a close analog with on-shell diagrams in ABJM
theory and the positive null Grassmannian \cite{Huang:2013owa}, so
it is natural to expect an analog of the amplituhedron for ABJM as
well. Should we expect any of the ideas in this paper to extend to
other field theories, with less or no supersymmetry, and beyond the
planar limit? As explained in \cite{N6}, the connection between
on-shell diagrams and the Grassmannian is valid for any theory in
four dimensions, reflecting only the building-up of more complicated
on-shell processes from gluing together the basic three-particle
amplitudes. The connection with the positive Grassmannian in
particular is universal for any planar theory: only the measure on
the Grassmannian determining the on-shell form differs from theory
to theory. Furthermore, on-shell BCFW representations of scattering
amplitudes are also widely available--at loop level for planar gauge
theories, and at the very least for gravitational tree amplitudes
(where there has been much recent progress from other points of view
\cite{Hodges:2012ym,Cachazo:2012da,Cachazo:2012kg,Skinner:2013xp,Cachazo:2013hca,Cachazo:2013iea}).
As already mentioned, one of the crucial clues leading to the
amplituhedron was the myriad of different BCFW representation of
tree amplitudes, with equivalences guaranteed by remarkable rational
function identities relating BCFW terms. We have finally come to
understand these representations and identities as simple
reflections of amplituhedron geometry. As we move beyond planar
${\cal N} = 4$ SYM, we encounter even {\it more} identities with
this character, such as the BCJ relations \cite{Bern:2008qj,
Bern:2010ue}. Indeed even sticking to planar ${\cal N} = 4$ SYM,
such identities, of a fundamentally non-planar origin, give rise to
remarkable relations between amplitudes with different cyclic
orderings of the external data. It is hard to believe that these
on-shell objects and the identities they satisfy only have a
geometric ``triangulation" interpretation in the planar case, while
the even richer structure beyond the planar limit have no geometric
interpretation at all. This provides a strong impetus to search for
a geometry underlying more general theories.

Planar ${\cal N} = 4$ SYM amplitudes are Yangian invariant, a fact
that is invisible in the  conventional field-theoretic description
in terms of amplitudes in one space or Wilson loops in the dual
space. We have become accustomed to such striking facts  in string
theory, which has a rich spectrum of $U$ dualities, that are
impossible to make manifest simultaneously in conventional string
perturbation theory. Indeed the Yangian symmetry of planar ${\cal N}
= 4$ SYM is just fermionic $T$-duality \cite{Berkovits:2008ic}.
The amplituhedron has now given us a new description of planar
${\cal N} = 4$ SYM amplitudes which does not have a usual
space-time/quantum mechanical description, but {\it does} make all
the symmetries manifest. This is not a ``duality" in the usual sense, since we are not identifying an equivalence
between existing theories with familiar physical interpretations. We
are seeing something rather different: new mathematical
structures for representing the physics without reference to
standard physical ideas, but with all symmetries manifest. Might
there be an analogous story for superstring scattering amplitudes?

\newpage

{\Large \bf Acknowledgements} \vskip .1in

We thank Zvi Bern, Jake Bourjaily, Freddy Cachazo, Simon Caron-Huot,
Clifford Cheung, Pierre Deligne, Lance Dixon, James Drummond, Sasha
Goncharov, Song He, Johannes Henn, Andrew Hodges, Yu-tin Huang,
Jared Kaplan, Gregory Korchemsky, David Kosower, Bob MacPherson,
Juan Maldacena, Lionel Mason, David McGady, Jan Plefka, Alex
Postnikov, Amit Sever, Dave Skinner, Mark Spradlin, Matthias
Staudacher, Hugh Thomas, Pedro Vieira, Anastasia Volovich, Lauren
Williams and Edward Witten for valuable discussions. Our research in
this area over the past many years owes an enormous debt of
gratitude to Edward Witten, Andrew Hodges, and especially Freddy
Cachazo and Jake Bourjaily, without whom this work would not have
been possible. N. A.-H. is supported by the Department of Energy
under grant number DE-FG02-91ER40654. J.T. is supported in part by
the David and Ellen Lee Postdoctoral Scholarship and by DOE grant
DE-FG03-92-ER40701 and also by NSF grant PHY-0756966.

\end{document}